\documentclass[twocolumn, titlepage,showpacs,preprintnumbers,amsmath,
amssymb,floatfix,nofootinbib]{revtex4}

\usepackage{graphicx}% Include figure files
\usepackage{dcolumn}% Align table columns on decimal point
\usepackage{bm}% bold math
\usepackage{amsxtra}
\usepackage{latexsym}
\usepackage{comment}

\newcommand{\ket}[1]{\mbox{\ensuremath{ | #1 \rangle }}}                           

\newcommand{\inprod}[2]{\mbox{\ensuremath{ \langle #1 | #2 \rangle}}}

\newcommand{\cop}[1]{\mbox{\ensuremath{ a_{#1}^{\dagger} }}}
\newcommand{\aop}[1]{\mbox{\ensuremath{ a_{#1}^{\phantom{\dagger} }}}}
\newcommand{\op}{\mbox{\ensuremath{P_{\infty}  }}}
\newcommand{\sline}{\mbox{\protect\rule[2pt]{0.4cm}{0.1pt}}}
\newcommand{\dashline}{\mbox{\protect\rule[2pt]{0.1cm}{0.1pt} \hspace{-4pt} 
    \protect\rule[2pt]{0.1cm}{0.1pt} \hspace{-4pt} \protect\rule[2pt]{0.1cm}{0.1pt}}}
\newcommand{\ddashline}{\mbox{$\cdot \cdot$ \hspace{-3pt}
    \protect\rule[2.1pt]{0.175cm}{0.6pt} \hspace{-3pt}
    $\cdot \cdot$}}

\newcommand{\floor}[1]{\mbox{\ensuremath{ \lfloor #1 \rfloor }}}
\newcommand{\ceil}[1]{\mbox{\ensuremath{ \lceil #1 \rceil }}}

\begin{document}

\title{Supercritical series expansion for the contact process in heterogeneous
  and disordered environments}

\author{C.~J.~ Neugebauer}  
 \email{cjn24@cam.ac.uk}  
\affiliation{Department of Chemistry, University of Cambridge,  
             Cambridge, UK}

\author{S.~N.~Taraskin}  
 \email{snt1000@cam.ac.uk}  
\affiliation{St. Catharine's College and Department of Chemistry, University of Cambridge,  
             Cambridge, UK}  

\date{\today}

\begin{abstract}

The supercritical series expansion of the survival probability for the
one-dimensional contact process in heterogeneous and disordered lattices is
used for the evaluation of the loci of critical points and critical exponents
$\beta$.
The heterogeneity and disorder are modeled by considering binary regular and
irregular lattices of nodes characterized by different recovery rates and
identical transmission rates.
Two analytical approaches based on \emph{Nested Pad\'e
  approximants} and \emph{Partial Differential approximants}    
were used in the case of expansions
with respect to two variables (two recovery rates) 
for the evaluation of the critical values and critical exponents. 
The critical exponents in heterogeneous systems are very close to those 
for the homogeneous contact process thus confirming that the contact process in
periodic heterogeneous environment belongs to the directed percolation
universality class. The disordered systems, in contrast, seem to have
continuously varying critical exponents.
\end{abstract}


\maketitle

\section{Introduction \label{sec:intro}}

In nonequilibrium statistical mechanics, phase transitions have been
identified and studied for some time now.
Similar to equilibrium systems, these phase transitions fall into a number
of different universality classes, one of which is the directed percolation
(DP) universality class~\cite{odor_04, hinrichsen_00}.
According to a conjecture by Janssen~\cite{janssen_81} and
Grassberger~\cite{grassberger_82}, all absorbing state phase
transitions with a scalar order parameter and no additional
conservation laws are characterized by DP critical
exponents.

One of the questions that has not been answered conclusively is how the
introduction of quenched disorder affects the universal critical behavior of
the DP class.
According to the Harris criterion~\cite{harris_74_2}, the critical exponents
should change even for weak disorder.
This has been investigated in particular for the contact process
(CP)~\cite{liggett_85} which is one of the archetypical models of the DP
universality class.
Currently, there are two alternative scenarios how the exponents change with
introduction of disorder: according to results in Refs.~\cite{moreira_96,
  cafiero_98, dickman_98, hooyberghs_03, hooyberghs_04}, they change continuously with degree
of disorder,
while Vojta~\cite{vojta_05} has suggested an abrupt change to the values in
the strong disorder limit corresponding to the universality class of the
random transverse Ising model. 
  
The above controversy can be addressed either by numerical or analytical
methods.
Among analytical methods, time-dependent perturbation theory as introduced by
Dickman and Jensen~\cite{dickman_91, jensen_93} for homogeneous $1d$
systems gives the most accurate numerical values. 
Technically, this has been done by using one-variable Pad\'e
approximants in the analysis of the series for the survival probability. 
In this paper, we extend their approach to heterogeneous and disordered  
$1d$ lattices and introduce the use of a Pad\'e approximants similar to 
the Nested Pad\'e approximants (NPA's)
\cite{guillaume_97, guillaume_00} in order to deal with two control 
parameters (two recovery rates), characteristic of binary lattices. 
A different approach based on partial differential approximants (PDA's) has
been used by Dantas and Stilck in Ref.~\cite{dantas_05}, who applied 
the supercritical series expansion to study the crossover between the $1d$ 
CP and the voter model~\cite{liggett_85},
thereby introducing a second control parameter to the perturbation
theory. 
We also use PDA's in order to compare results from the two
extrapolation methods.

The aim of the paper is two-fold: first, we present and discuss 
the technical details of
the supercritical series expansions in the case of two variables, i.e.\ of two
different recovery rates
characteristic of nodes of two different types, which are
different from the well-studied one-parameter case~\cite{dickman_91,
  jensen_93}.
Second, we investigate the range of applicability and the effects of
variations of our procedure based on NPA's  
on the estimates of critical values and
exponents and compare these results to those obtained by employing PDA's. 

The structure of the paper is as follows: 
we introduce the CP in Sec.~\ref{sec:model}. 
The supercritical series expansion, 
 the analysis of the resulting two-variable series and the
configurational averaging of the order parameter for
disordered environments are discussed in Sec.~\ref{sec:analysis}.  
In Sec.~\ref{sec:results}, we present the results in terms of phase diagrams
and the critical exponent $\beta$ for the different systems, which we then
discuss in the last Sec.~\ref{sec:DiscConc}.

\section{Model \label{sec:model}}

The CP, originally introduced by Harris~\cite{harris_74}, is
a spatial SIS (Susceptible-Infected-Susceptible)-model for the spread of
epidemics through a network.
In a network, usually taken to be
a hyper-cubic lattice, the nodes or sites can be in one of two
states: ``infected'' or ``susceptible''. 
A susceptible/vacant site
$i$ can be infected by a neighboring infected/occupied site $j$,
while an occupied site $k$ can spontaneously  
recover and become susceptible again. 
These processes occur with rates
$\lambda_{ij}/z$ and $\mu_k$, respectively, where $z$ is the number of nearest
neighbors in the lattice.
 
In all dimensions, the CP undergoes a nonequilibrium phase transition into an absorbing state
which does not allow any further time evolution. 
For a fixed set of transmission rates, this occurs when the recovery rates 
become smaller than a set of critical values.
At this point, the survival probability, that is the probability that process
will not enter the absorbing state (in the thermodynamic limit),  
becomes greater than zero. 
Several observables can be identified that describe
the critical
state of the CP, such as the average density of infected sites, $\rho\,(t)$,
or the survival probability up to time $t$ after starting from a single
seed, $P_{\text{s}}\,(t)$. 
As $t \to \infty$, 
in a homogeneous system ($\lambda_{ij} = 1$ and $\mu_k = \mu$),
close to the critical point these
quantities are expected to behave like \cite{hinrichsen_00}
\begin{eqnarray}
  \lim_{t \to \infty}\rho (t) \equiv \rho_{\text{stat}} \sim  \Delta^{\beta}  \qquad 
  \lim_{t \to \infty} P_{\text{s}} (t) \equiv P_{\infty} \sim  \Delta^{\beta}. 
\end{eqnarray} 
Here $\beta$ is the critical exponent associated with stationary behavior of the
order parameter and $\Delta = \mu_c - \mu > 0$, where $\mu_c$ is the critical
point for the control parameter recovery rate $\mu$.

Each site $i$ of the system of $N_{\text{s}}$ nodes obeying the rules of the CP can be in two states:
$\ket{\sigma_i}$ where $\sigma_i= 0$ or $\sigma_i=1$
so that a microstate of the system with $N_{\text{s}}$ sites can be written as
\begin{eqnarray*}
\ket{\sigma} =\ket{\sigma_1} \otimes \ldots \otimes
\ket{\sigma_{N_{\text{s}}}} = \bigotimes_i^{N_{\text{s}}} \ket{\sigma_i} = \bigotimes_i^{N_{\text{s}}} 
\left( 
\begin{array}{c}
1-\sigma_i  \\ 
\sigma_i
\end{array}
\right)  
\end{eqnarray*}
This representation ensures that the microstate vectors form a
 $2^{N_{\text{s}}}$-dimensional orthonormal basis. The state of the system at time $t$ is
\begin{eqnarray}
\ket{P(t)}=\sum_{\sigma}P (\sigma,t)\,\ket{\sigma}
\end{eqnarray}
where $P (\sigma,t) = \inprod{\sigma}{P(t)}$ is the probability that the system is found in
 microstate $\ket{\sigma}$.
 The time-evolution of the state of the system is governed by the master
 equation
\begin{eqnarray}
\partial_t \ket{P(t)} = \hat{L} \ket{P(t)} \label{eq:lvmaster}
\end{eqnarray}
where $\hat{L}$ is the Liouville operator whose non-zero off-diagonal elements in this basis are 
the transition rates between microstates that for the CP differ in their
occupation number by one. 
This operator describes the probability flow between
different microstates and is thus represented by a stochastic matrix in which
all the diagonal elements are the sums of the off-diagonal elements in the 
corresponding columns taken with the opposite sign.
  
In a formalism introduced by Doi~\cite{doi_76} and
Peliti~\cite{peliti_85} for stochastic systems, ``annihilation'' and ``creation'' operators
on site $i$, $\aop{i}$ and $\cop{i}$, respectively, are defined, such that
$\aop{i}\, \ket{\sigma_i} = \sigma_i \,\ket{\sigma_i-1}$ and $\cop{i}
\,\ket{\sigma_i}=(1-\sigma_i)\, \ket{\sigma_i+1}$ 
(e.g.\ $\aop{i} \ket{0} = 0$) and that they obey hard-core bosonic commutation relations. 

For simplicity, we assume all the transmission rates to be the same and 
define the time scale by setting $\lambda_{ij}=1$. 
The recovery rates for binary systems are 
characterized by one of two values $\mu_i=\{\mu_A, \mu_B\}$. 
For such models, the operator $\hat{L}$ in the one-dimensional case
reads as
\begin{eqnarray}
\hat{L}=\mu\,\hat{W} + \hat{V}~, 
\end{eqnarray}
where the operators $\mu\,\hat{W}$ and $\hat{V}$ are 
\begin{eqnarray}
\mu\,\hat{W} &=& \mu \sum_i \frac{\mu_i}{\mu}\, (1-\cop{i})\,\aop{i}
\label{eq:def_W}~, \\
\hat{V} &=& \sum_i \frac{1}{2}\,(1-\aop{i})\,\cop{i}\,
(\cop{i-1}\aop{i-1}+\cop{i+1}\aop{i+1}) \label{eq:def_V}~.
\end{eqnarray}
The parameter $\mu$ is introduced for convenience 
and $\mu_i/\mu \sim 1$ with both recovery rates being close to the homogeneous
critical point and thus $\mu_i \alt 1$.
The operator $\hat{V}$ creates offspring at the nearest neighbors of an occupied
 sites, while $\hat{W}$ destroys occupation at sites.
The above formalism is useful for the supercritical series expansion in $\mu$
described below.

%%%%%%%%%%%%%%%%%%%%%%%%%%%%%%%%%%%%%%%%%%%%%%%%%%%%%%%%%%%%%%%%%%%%%%%%%%%%%%

\section{Analysis \label{sec:analysis}}

\subsection{Supercritical Series Expansion\label{sec:series_expansion}}

The supercritical series expansion is a perturbation series for the ultimate
survival probability $P_{\infty}$
which is taken to be the order parameter, with $P_{\infty} > 0$ being
characteristic of the CP in the active phase. 
To probe the long-time limit of the system, the Laplace transform of the
probability state vector is taken,
\begin{eqnarray}
\ket{\tilde{P}\,(s)} = \int_0^{\infty} dt\, e^{-st}\, \ket{P\,(t)}\label{eq:lt}~,
\end{eqnarray}
so that a standard identity of Laplace-transform theory, $\lim_{t \rightarrow \infty} f(t) = \lim_{s
  \rightarrow 0} s\,\tilde{f}(s)$, can be used.
The ultimate survival probability is then given by
\begin{eqnarray}
P_{\infty}=\lim_{s \rightarrow 0} \left(1 -
  s\,\inprod{0}{\tilde{P}\,(s)}\right) \label{eq:sp}~.
\end{eqnarray}
Inserting the formal solution
$\ket{P\,(t)}=e^{-\hat{L}\,t}\ket{P\,(0)}$ of Eq.~(\ref{eq:lvmaster}) into
Eq.~(\ref{eq:lt}) results in
\begin{eqnarray}
\ket{\tilde{P}\,(s)} = 
(s-\mu\,\hat{W}-\hat{V})^{-1} \ket{P\,(0)} \label{e2}~.
\end{eqnarray}
We can then formally expand the operator on the right hand-side of Eq.~(\ref{e2}) 
in a Taylor series with respect to the small parameter
$\mu$ thus obtaining the following supercritical expansion
\begin{eqnarray}
\ket{\tilde{P}\,(s)}& =& \ket{\widetilde{P}_0(s)}+ \mu \ket{\widetilde{P}_1(s)} + \mu^2
  \ket{\widetilde{P}_2(s)} + \ldots
\end{eqnarray}
where the vectors $\ket{\widetilde{P}_{n}(s)}$ obey the following recursion relations:
\begin{eqnarray}
\ket{\widetilde{P}_0(s)}&=&(s-\hat{V})^{-1}\ket{P(0)}\\
\ket{\widetilde{P}_n(s)}&=&(s-\hat{V})^{-1} \hat{W}
\ket{\widetilde{P}_{n-1}(s)} 
\quad \mbox{for}~n \geq 1~.
\label{e3}
\end{eqnarray}
The action of the operator $\hat{W}$ on a given configuration can 
be straightforwardly computed using
its definition given by Eq.~(\ref{eq:def_W}), i.e.\ 
\begin{eqnarray}
  \hat{W} \ket{\sigma} = \sum_i^m \frac{\mu_i}{\mu} \left( \ket{\overline{\sigma}^i}
  - \ket{\sigma} \right).
\end{eqnarray}
The summation is taken over all $m$ occupied sites in state $\sigma$ and
$\ket{\overline{\sigma}^i}=\aop{i}\ket{\sigma}$, i.e.\ the
$\overline{\sigma}^i$ and $\sigma$ have the same occupation except for site $i$
being occupied in state $\sigma$ and vacant in $\overline{\sigma}^i$.

The action of the operator $(s-\hat{V})^{-1}$ on a given configuration,
$\ket{\sigma}$, can be computed with the use
of the following identity, 
\begin{eqnarray}
(s-\hat{V})^{-1}=s^{-1} + s^{-1}(s-\hat{V})^{-1}\hat{V}~, 
\label{eq:identity}
\end{eqnarray}
so that 
\begin{widetext}
  \begin{eqnarray}
    (s-\hat{V})^{-1} \ket{\sigma} = \frac{1}{s+q_1/2+q_2} 
    \left(\ket{\sigma} + (s-\hat{V})^{-1} \left[\frac{1}{2} 
      \sum_i^{q_1} \ket{\tilde{\sigma}^{1,i}} + 
      \sum_i^{q_2} \ket{\tilde{\sigma}^{2,i}}\right] \right)
    \label{eq:q1q2}
  \end{eqnarray}
\end{widetext}
where the sums represent the action of the operator $\hat{V}$ on the state
$\sigma$ with vacant sites of two types: sites which have one (first sum in
Eq.~(\ref{eq:q1q2}) is taken over a number $q_1$ of such sites) or two (second
sum in Eq.~(\ref{eq:q1q2}) is taken over $q_2$ of such sites) occupied nearest
neighbors. 
The vectors $\ket{\overline{\sigma}^{1,i}}$ and 
$\ket{\overline{\sigma}^{2,i}}$ represent the states in which the formerly
vacant sites $i$ of the first and second type, respectively, are now
occupied. 
To go further we should use the recursive nature of the operator  
$(s-\hat{V})^{-1}$ and substitute its representation given by 
Eq.~(\ref{eq:identity}) into Eq.~(\ref{eq:q1q2}). 
Such a procedure can generate an infinite number of new configurations. 
However, when we calculate the survival probability perturbatively up to 
a given order $N$ in $\mu$ for the initial
condition of a single occupied site, it is only necessary to retain states
with up to $N$ occupied sites. 
This is due the fact that the
annihilation operator $\hat{W}$ in an expansion up to order $\mu^N$ acts $N$
times on any generated state, after which remaining states will be projected
onto the absorbing state, thereby projecting out any states with more than $N$
occupied sites.
Following this procedure, we can perturbatively calculate the survival
probability $\op$ and thus find the critical point where the survival
probability becomes zero.
In order to obtain good numerical estimates of this critical value and compute
critical exponents, it is necessary to employ numerical methods such as Pad\'e
approximants~\cite{baker_96}. 

%%%%%%%%%%%%%%%%%%%%%%%%%%%%%%%%%%%%%%%%%%%%%%%%%%%%%%%%%%%%%%%%%%%%%%%%%

\subsection{Nested Pad\'e Approximants: Critical Values and Exponent\label{sec:pade}}

For systems with two different recovery rates, $\mu_A$ and $\mu_B$,
the survival probability (see Eq.~(\ref{eq:sp})) expanded in series is a
polynomial in these two variables,
\begin{eqnarray}
\op (\mu_A,\mu_B)=1 - \sum_{n=1}^N \sum_{m=0}^{n} ~c_{nm} ~\mu_B^m
~\mu_A^{n-m} ~.
\end{eqnarray}
The critical line $\mu_B^{(\text{c})}=\mu_B^{(\text{c})}(\mu_A^{(\text{c})})$ that separates 
the absorbing state from the active
state is a solution to the equation $\op (\mu_A, \mu_B)=0$ corresponding to
the smallest (real) root.
In practice, just finding the roots of the polynomial -- the truncation of an
infinite series at finite order -- does not produce very
good estimates of the critical values. 
Better results are obtained by using d-log Pad\'e approximants
\cite{guttmann_89}: given an expansion in one variable, $\op(\mu)$, up to
order $N$, the Pad\'e approximant
\cite{bender_78, baker_96} of the a series for $\partial_{\mu} \ln \op(\mu)$
is formed.
Technically, this is done by expanding the denominator of $\partial_{\mu} \ln
\op(\mu) = \partial_{\mu} P(\mu) / P(\mu)$ up to order $N-1$ and thus
obtaining a polynomial $f_{N-1}(\mu)$ for this fraction, the Pad\'e
approximant of which is then constructed.
The first positive (real) pole and its residue then provide good estimates of
the critical value and the critical exponent of $\op$, respectively.  
The extension of this approach to two variables,
however, is not straightforward -- there is a number of different
multi-variable generalizations \cite{chisholm_73, fisher_77} of the
one-variable Pad\'e approximation.
In this work, we employ a scheme similar to the NPA's~\cite{guillaume_97,
  guillaume_00} in which we in turn form 
one-variable Pad\'e approximants with respect to the two variables.

To this end, 
we transform the variables $(\mu_A, \mu_B)$ to the following three 
 more convenient sets, T1, T2 and T3. 
The first transformation, T1, is symmetric so that the values 
$(\mu_A, \mu_B)$ are replaced by $(\bar{\mu}, \delta)$ where 
$\bar{\mu}=(\mu_A + \mu_B)/2$ and $\delta=(\mu_A - \mu_B)/2$. 
The transformations T2 and T3 are asymmetric with $(\mu_A, \mu_B)$ 
replaced by either $(\bar{\mu}=\mu_A, \delta=\mu_A - \mu_B)$ or 
 $(\bar{\mu}=\mu_B, \delta=\mu_B - \mu_A)$, respectively. 
Expanding $\frac{\partial}{\partial\bar{\mu}} \ln \op(\bar{\mu},
\delta)=\partial_{\bar{\mu}} \op / \op(\bar{\mu},\delta)$ up to order $N-1$ in
$\bar{\mu}$ and $\delta$, we obtain
%i.e.\ 
%
\begin{eqnarray}
\frac{\partial}{\partial\bar{\mu}} \ln \op(\bar{\mu},
\delta) = \sum_{n=0}^{N-1} f_n (\delta) \bar{\mu}^n + O(\bar{\mu}^N).
\end{eqnarray}
We then form the Pad\'e approximants of the coefficients,
\begin{eqnarray} 
f_n (\delta) = \frac{\sum_{m=0}^{J} q_m \delta^m}
{1+\sum_{m=1}^{K} q_{J+m} \delta^m},
\end{eqnarray}
where $N-1-n=J+K$.
As always with Pad\'e approximants, we have the freedom to choose $J$ and $K$
for a given $n$.
For even $N-1-n$, we use diagonal Pad\'e approximants $[J, J]$ with
$J=K=(N-1-n)/2$, while for odd $N-1-n$, we form approximants $[J-1, J]$ with
$J=(N-n)/2$. 

Then, in turn, we form the Pad\'e approximant with respect to $\bar{\mu}$,   
\begin{eqnarray}
\partial_{\bar{\mu}} \ln \op = \frac{\sum_{n=0}^{L} r_n(\delta)
  \,\bar{\mu}^n}{1+\sum_{n=1}^{M} r_{L+n} (\delta)\, \bar{\mu}^n} + O(\bar{\mu}^{N}) \label{eq:npade1}
\end{eqnarray}
where $N-1=L+M$.
%
%%%%%%%%%%%%%%%%%%%%%%%%%%%%%%%%%
A more graphical representation of the scheme is the following:
\begin{widetext}
  \begin{eqnarray}
    \partial_{\bar{\mu}} \ln \op &=& \underbrace{(a_{0,0} + a_{0,1} \delta + \ldots + a_{0,N-1}\,
    \delta^{N-1})} + \underbrace{(a_{1,0} + \ldots a_{1,N-1}\,\delta^{N-2})} \bar{\mu} + \ldots
    a_{N-1,0}\, \bar{\mu}^{N-1}  + O(\bar{\mu}^N) \label{eq:step1}\\
    &=& \underbrace{\quad\big[\floor{(N-1)/2}, \ceil{(N-1)/2}\big]_\delta \hspace{13pt} + \big[\floor{(N-3)/2},
    \ceil{(N-1)/2}\big]_\delta\,\bar{\mu} +  \ldots  a_{N-1,0}\, \bar{\mu}^{N-1}} +
    O(\bar{\mu}^N) \label{eq:step2}\\
    &=& \hspace{110pt} \big[\floor{(N-1)/2},\ceil{(N-1)/2}\big]_{\bar{\mu}} + O(\bar{\mu}^N) \label{eq:step3}
  \end{eqnarray}
\end{widetext}
%%%%%%%%%%%%%%%%%%%%%%%%%%%%%%%%%
%
Here we have denoted the formation of the Pad\'e approximant with respect to a
variable $x$ with numerator degree $N$ and denominator degree $M$ as
$\big[N,M\big]_x$ ($\floor{\cdot}$ and $\ceil{\cdot}$ are the floor and ceiling functions, respectively).

Thus for any given $\delta$, we can find the corresponding
pole of $\partial_{\bar{\mu}} \ln \op(\bar{\mu}, \delta)$, which is then taken
to as the
critical value
$\bar{\mu}^{(\text{c})}(\delta)$, yielding a point on the critical line,
$(\mu_A^{(\text{c})}= \bar{\mu}^{(\text{c})},
\mu_B^{(\text{c})}= \bar{\mu}^{(\text{c})}+\delta)$, $(\mu_A^{(\text{c})} = \bar{\mu}^{(\text{c})}+\delta,
\mu_B^{(\text{c})} = \bar{\mu}^{(\text{c})})$ or
$(\mu_A^{(\text{c})} = \bar{\mu}^{(\text{c})} + \delta, \mu_B^{(\text{c})} = 
\bar{\mu}^{(\text{c})} - \delta)$, depending on the initial transformation.

It turns out that occasionally the first positive real roots are unphysical
ones that appear before the physical solution.
However, these roots are very closely matched by roots of the numerator of
the Pad\'e approximant of $\partial_{\bar{\mu}} \ln \op(\mu_A, \delta)$, so
that these two cancel each 
other, leaving the physical root as the solution.
In order to extract unphysical roots a further parameter $\gamma$ has been
introduced. 
Two roots, $x_1$ and $x_2$, of the
numerator, $n(x)$, and the denominator, $d(x)$, respectively, i.e.\ 
$(x-x_1) n(x)/\left((x-x_2) d(x)\right)$,
are considered to be the same value and cancel if $|x_1 - x_2| < \gamma$.
All results presented below are obtained by setting $\gamma = 10^{-3}$.

In order to evaluate the stability of a certain pole, several
approximants are formed with respect to $\bar{\mu}$ close to the diagonal
approximant, 
e.g.\ for even $N-1= 2 K$, we compute $[K,K], [K-1,K], [K, K-1], \ldots [K-2,
K-2]$ and take the average over these poles. 
Once we obtained the critical value, the critical exponent $\beta$ associated
with the order parameter $\op$ can be found
as well, as it is just the residue at the pole $\bar{\mu}^{(\text{c})}$. 
Again, the average
over the residues for different Pad\'e approximants is taken. 
As the error
that we could
extract from employing this method is small and does not take into
account the inherent error in the series expansion, we perform this averaging
to minimize the effects due to a particular choice of approximant and use the
standard deviation only to evaluate the numerical stability of a pole.

\subsection{Partial Differential Approximants: Critical Values\label{sec:pda}}

Another method for estimating critical values given a finite
two-variable series are the PDA's
originally developed by Fisher~\cite{fisher_77-2} in order to investigate
multicritical points.

The starting point of this method is that one is given a finite
two-variable polynomial, 
$F(x,y) = \sum_{(i,j) \in \mathbf{S}} f_{ij} x^i y^j$,
that approximates a
true function $f(x,y)$, that is expected to have the following form,
\begin{eqnarray}
  f(x,y) = ((x - x_c) + \alpha (y - y_c))^\beta \phi(x,y)
\end{eqnarray}
where $\phi (x,y)$ is some general function with $\phi(x_c,y_c) \neq
0$.
The set $\mathbf{S}$ is a so-called label set that contains the pairs of
powers $(i,j)$ of $x$ and $y$ only if $f_{ij} \neq 0$, i.e.\ 
$(i,j) \in \mathbf{S}$ if $F(x,y)$ has a non-zero term of order $x^i y^j$.
For fixed label sets $\mathbf{L}$, $\mathbf{M}$ and $\mathbf{N}$, it is possible to find 
%the coefficients $p_{ij}$, $q_{ij}$ and $r_{ij}$ of the 
polynomials $P_{\mathbf{L}}(x,y) = \sum_{(i,j) \in \mathbf{L}} p_{ij}
x^i y^j$, $Q_{\mathbf{M}}(x,y)= \sum_{(i,j) \in \mathbf{M}} q_{ij} x^i y^j$ and
$R_{\mathbf{N}}(x,y) = \sum_{(i,j) \in \mathbf{N}} r_{ij} x^i y^j$ such that they
satisfy the \emph{defining} equation
\begin{eqnarray}
  P_{\mathbf{L}}(x,y) F = Q_{\mathbf{M}}(x,y) \frac{\partial F}{\partial x} + R_{\mathbf{N}}(x,y)\frac{\partial
    F}{\partial y} + E_{\mathbf{K}}(x,y)
  \label{eq:PDA}
\end{eqnarray}
with $E_{\mathbf{K}} = \sum_{(i,j) \notin \mathbf{K}} e_{ij} x^i y^j$
denoting a sum of non-zero terms whose powers are not in the \emph{matching
  set} $ \mathbf{K}$.
This matching set defines for which powers $x^i y^j$
Eq.~(\ref{eq:PDA}) should hold exactly with $e_{ij} = 0$, while for
$(m,n) \notin \mathbf{K}$, the values of $e_{mn}$ are allowed to be non-zero.
In order for Eq.~(\ref{eq:PDA}) to be a solvable set of linear equations,
the label sets must obey the constraint that 
the label sets $\mathbf{L}$, $\mathbf{M}$ and $\mathbf{N}$ must together contain
one more element than the matching set $\mathbf{K}$
because of the conventional choice of $p_{00} = 1$.  

Once the polynomials $P_{\mathbf{L}}$, $Q_{\mathbf{M}}$ and
$R_{\mathbf{N}}$ are found, 
e.g. by using an algorithm
proposed by Styer~\cite{styer_90}, they can be used to find an
estimate for the line of critical points 
by the method of characteristics (e.g. see~\cite{stilck_81}). 
According to this method, consider a single curve of points which only depends on a single
parameter $\tau$, $\mathbf{x}(\tau) = (x(\tau), y(\tau))$.
The rate of change of $F(x(\tau), y(\tau))$ along this line is
$\mathrm{d}F / \mathrm{d}\tau  = (\partial F/\partial x) (\mathrm{d}x
/ \mathrm{d}\tau) + (\partial F/\partial y) (\mathrm{d}y
/ \mathrm{d}\tau)$.
The survival probability $P_\infty$, for which we are going to apply
this method, is zero along the critical line.
Thus, considering the case where $F(x(\tau), y(\tau)) =
0$, it can be seen that the curve described by the equations
\begin{eqnarray}
\frac{\mathrm{d}x}{\mathrm{d}\tau} &=& Q_{\mathbf{M}} (x(\tau), y(\tau))\\
\frac{\mathrm{d}y}{\mathrm{d}\tau} &=& R_{\mathbf{N}} (x(\tau), y(\tau))
\end{eqnarray}
and substituted in Eq.~(\ref{eq:PDA}) yields 
the relation $0 = (\partial F/\partial x) (\mathrm{d}x/\mathrm{d}\tau) + (\partial
F/\partial y) (\mathrm{d}y / \mathrm{d}\tau)$.
Together with a suitable initial condition, this curve is therefore equivalent to
the critical line as $F$ does not change along the curve $\mathbf{x}(\tau)$.
This suitable initial condition has to be a known point on the
critical line: in our case, this 
is the critical point of the
homogeneous system at which $x = y = x_c$.

\subsection{Configurational Averaging \label{sec:configav}}

%The scheme above is 
The schemes above are straightforwardly applied to heterogeneous topologically
ordered systems.
In topologically disordered systems, the survival probability has to be
averaged over different realizations of disorder. 
For concreteness, we consider disorder only in recovery rates $\mu_i$  
on different nodes $i$ assuming that $\mu_i$ are independent random variables
distributed according to the probability distribution functions 
$\rho(\mu_i)$. 
A configurationally averaged survival probability is then given 
by the following  expression,
$\langle \op \rangle=\int \op(\mu_i) \prod_i\rho\,(\mu_i)\text{d}\mu_i $. 
For simplicity, we consider 
a bimodal distribution of recovery rates,
\begin{eqnarray}
  \rho\,(\mu_i) = p\, \delta(\mu_i - \mu_A) + (1-p)\, 
\delta(\mu_i - \mu_B)~, 
\label{eq:pdf}
\end{eqnarray}
i.e.\ the nodes $A$ (hosts), characterized by recovery rate $\mu_A$, 
of concentration $p$ and $B$ (impurities) of concentration $1-p$ are
randomly and independently placed on the sites of a regular chain.  

Using the series expansion of order $N$ for $\op$ is equivalent 
to considering the CP on the finite chain of length $2N-1$, %so that
i.e.\ for a given value of $n \le N$ all states on $2 n - 1$ sites with the
origin at 
its center with at most $n$ sites occupied contribute, 
so that
\begin{eqnarray}
 \langle \op
\rangle=1 + \sum_{n=1}^N \sum_{m=0}^{n}\,\langle c_{nm}\rangle \,\mu_B^m
\,\mu_A^{n-m}
\end{eqnarray}
where
\begin{eqnarray}
  \langle c_{nm} \rangle = \sum_{k=0}^{2 n - 1} \sum_{\{\mathbf{c}:
    \,N_{B}(\mathbf{c}) = k\}} (1-p)^k p^{2n-1-k} c_{nm} (\mathbf{c}) \label{eq:av} 
\end{eqnarray}
with the second sum running over all disorder configurations 
$\mathbf{c} = (\mu_{-n+1}, \ldots, \mu_0 , \ldots , \mu_{n-1})$ that have a
certain number $N_{B} (\mathbf{c})$ of impurity sites $B$. 
The values of $c_{nm}$ are the coefficients in the expansion of the survival
probability for a particular disorder configuration.
The factor $(1-p)^k p^{2n-1-k}$ stems from the fact that the probability of
having a particular disorder configuration is just the product of the
probabilities of any site being either $\mu_A$ or $\mu_B$, drawn from the
bimodal distribution given by Eq.~(\ref{eq:pdf}). 

The memory requirements in calculations of the coefficients in the series
expansions impose  a restriction, $N\le N_{\text{max}}$, 
on the highest order of expansion for disordered lattices, 
$N_{\text{max}} = 19$, 
which is rather lower than
for homogeneous~\cite{jensen_93} and heterogeneous cases, $N_{\text{max}} = 24$.  
The exact configurational averaging discussed above, exploiting symmetry 
about the origin and under the exchange of $\mu_A$
and $\mu_B$, is not possible for 
such high orders due to computational cost when dealing with a very 
large number (of $O(2^{2 N_{\text{max}}})$) of configurations. 
We have been able to undertake the exact configurational averaging up to 
order $N_{\text{c,max}}=12$. 

For higher orders, $N_{\text{c,max}}\le N \le N_{\text{max}}$, 
 a configurationally averaged survival
probability is calculated approximately by only including disorder 
configurations that have
no more than a certain number of impurities in the averaging.
Assuming that each coefficient is of the order of the
coefficient in a homogeneous system, $c_{nm} (\mathbf{c}) \sim c_n$, 
then each coefficient $\langle c_{nm} \rangle$ will remain of the same order
of magnitude if we choose a maximum number of impurities $k_{\text{max}}(n,p)$ in 
\begin{eqnarray}
  \langle c_{nm} \rangle \sim c_n \sum_{k=0}^{k_{\text{max}}} {2n-1 \choose k} (1-p)^k p^{2n-1-k} \label{eq:av2} 
\end{eqnarray}
such that the sum is close to unity.
For $(1-p) \ll 1$ and large $n \gg 1$, one can choose $k_{\text{max}} \ll 2 n - 1$, e.g.\
for $k_{\text{max}} (19, 0.96) = 2$, $\sum_{k=0}^{k_{\text{max}}} 
{2n-1 \choose k} (1-p)^k p^{2n-1-k} = 0.817$. 
For lower orders, the weight of the configurations that are dropped decreases
for constant $k_{\text{max}}$, but, because it becomes computationally
feasible, we choose to increase
$k_{\text{max}}$ by one for each lower order by letting $k_{\text{max}}(N_{\text{max}}-i,p)= i+2$.
This way, the lower the order the closer the approximate configurational
averaging is to the exact one.
With $k_{\text{max}}(N_{\text{max}}-i,p) = i+2$, the averaging becomes exact
from order $n \leq N/3 + 1$.

%%%%%%%%%%%%%%%%%%%%%%%%%%%%%%%%%%%%%%%%%%%%%%%%%%%%%%%%%%%%%%%%%%%%%%%%
%%%%%%%%%%%%%%%%   RESULTS  %%%%%%%%%%%%%%%%%%%%%%%%%%%%%%%%%%%%%%%%%%%
%%%%%%%%%%%%%%%%%%%%%%%%%%%%%%%%%%%%%%%%%%%%%%%%%%%%%%%%%%%%%%%%%%%%%%%%%

\section{Results\label{sec:results}}

In this section, we will present the results for critical values and critical
exponents obtained by the analysis described above and applied to different
systems.
These results come in the form of phase diagrams in which the
 critical points are plotted in the rate-space plane $(\mu_A, \mu_B)$ 
or as plots of the critical exponent $\beta$ as a function of the 
critical rate $\mu_A^{(\text{c})}$.

In Ref.~\cite{neugebauer_06}, it has been demonstrated that for the
$1d$ CP the line of
critical points close to the
homogeneous critical point, $(\mu_A = \mu_{\text{c}}, \mu_B = \mu_{\text{c}})$ is well described
by the relation
\begin{eqnarray}
\mu_{\text{c}} \simeq \exp \left( \mathrm{E}[\ln(\mu_i)] \right)~,
\label{eq:critline}
\end{eqnarray} 
where $\mathrm{E}[\cdot]$ denotes the expectation value with respect to the distribution of
the recovery rates $\mu_i$.
In this work, we will compare the critical points to the line given by
Eq.~(\ref{eq:critline}) to examine how far from the homogeneous critical point the
relationship describes the critical line well.

In all figures below, we only keep critical points and
critical exponents that have a standard deviations from the mean values of
less than $0.001$ and $0.005$, respectively, after
averaging over the Pad\'e approximants as described in Sec.~\ref{sec:pade}.

\subsection{Periodic Lattices}

%%%%%%%%%%%%%%%%%%%%%%%%%%%%%%%%%%%%%%%%%%%%%%%%%%%%%%%%%%%%%%%%%%%%%%%%%%%
\begin{figure}
  \begin{center}
    \scalebox{0.35}{\includegraphics[angle=0]{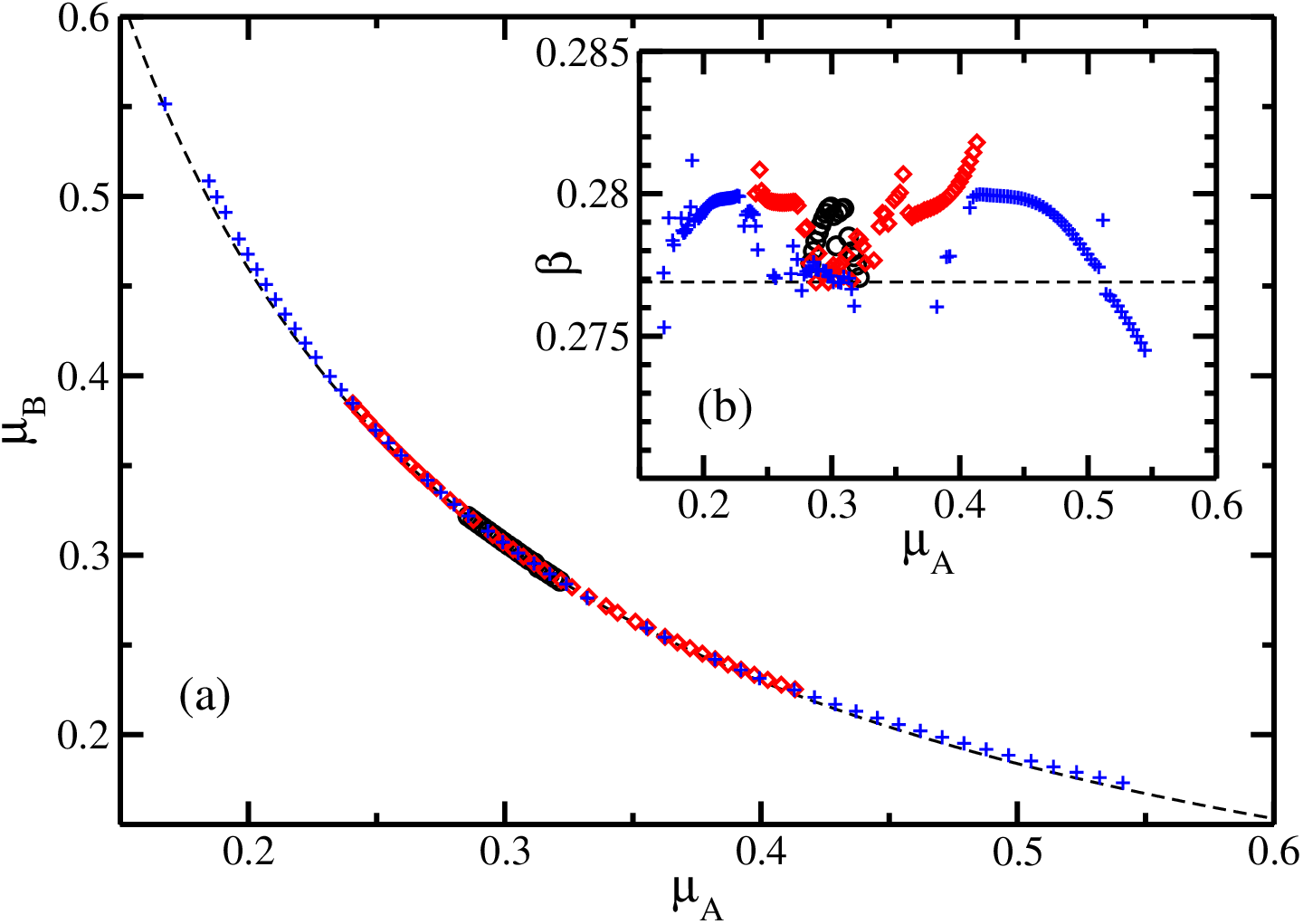}}
  \end{center}
  \caption{(Color online) Comparison of critical points (a) and critical
    exponents (b) of the $AB$ system obtained from series
    expansion for different orders
    $N$: $N = 10$ ($\circ$), $N=17$ ($\diamond$) and $N=24$ ($+$).
    The symmetric transformation T1 was used.
    The dashed lines (\dashline) are given by Eq.~(\ref{eq:critline}) in (a)
    and by $\beta$ at the homogeneous critical
    point, $\beta = 0.2769$, from series expansions~\cite{jensen_93} for the
    CP, in (b).
  } 
  \label{fig:comparison-AB}
\end{figure}
%%%%%%%%%%%%%%%%%%%%%%%%%%%%%%%%%%%%%%%%%%%%%%%%%%%%%%%%%%%%%%%%%%%%%%%%%%%%%%%%%%%%%%%%%
%
%%%%%%%%%%%%%%%%%%%%%%%%%%%%%%%%%%%%%%%%%%%%%%%%%%%%%%%%%%%%%%%%%%%%%%%%%%%
\begin{figure}
  \begin{center}
    \scalebox{0.35}{\includegraphics[angle=0]{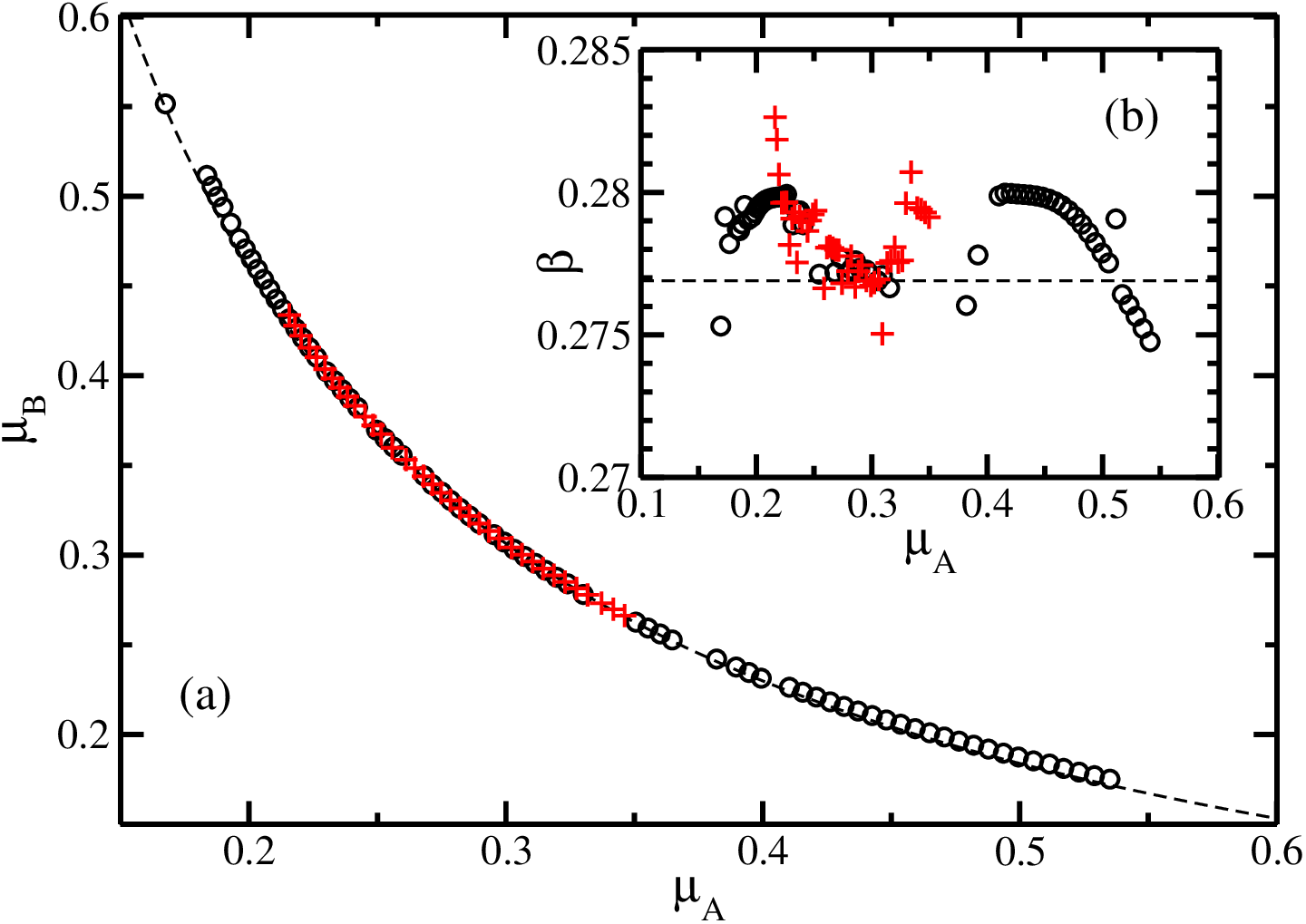}}
   \end{center}
  \caption{(Color online) Comparison of the critical points (a) and
    critical exponents (b) of the $AB$ system obtained from
    series expansions up to order $N=24$ with different transformations,
    T1 ($\circ$) and T2 ($+$).
    The dashed lines (\dashline) are given by Eq.~(\ref{eq:critline}) in (a)
    and by $\beta$ at the homogeneous critical
    point, $\beta = 0.2769$, from series expansions~\cite{jensen_93} for the
    CP, in (b).
  } 
  \label{fig:comparison-AB-T1+T2}
\end{figure}
%%%%%%%%%%%%%%%%%%%%%%%%%%%%%%%%%%%%%%%%%%%%%%%%%%%%%%%%%%%%%%%%%%%%%%%%%%%%%%%%%%%%%%%%%
%
%%%%%%%%%%%%%%%%%%%%%%%%%%%%%%%%%%%%%%%%%%%%%%%%%%%%%%%%%%%%%%%%%%%%%%%%%%%
\begin{figure}
  \begin{center}
    \scalebox{0.35}{\includegraphics[angle=0]{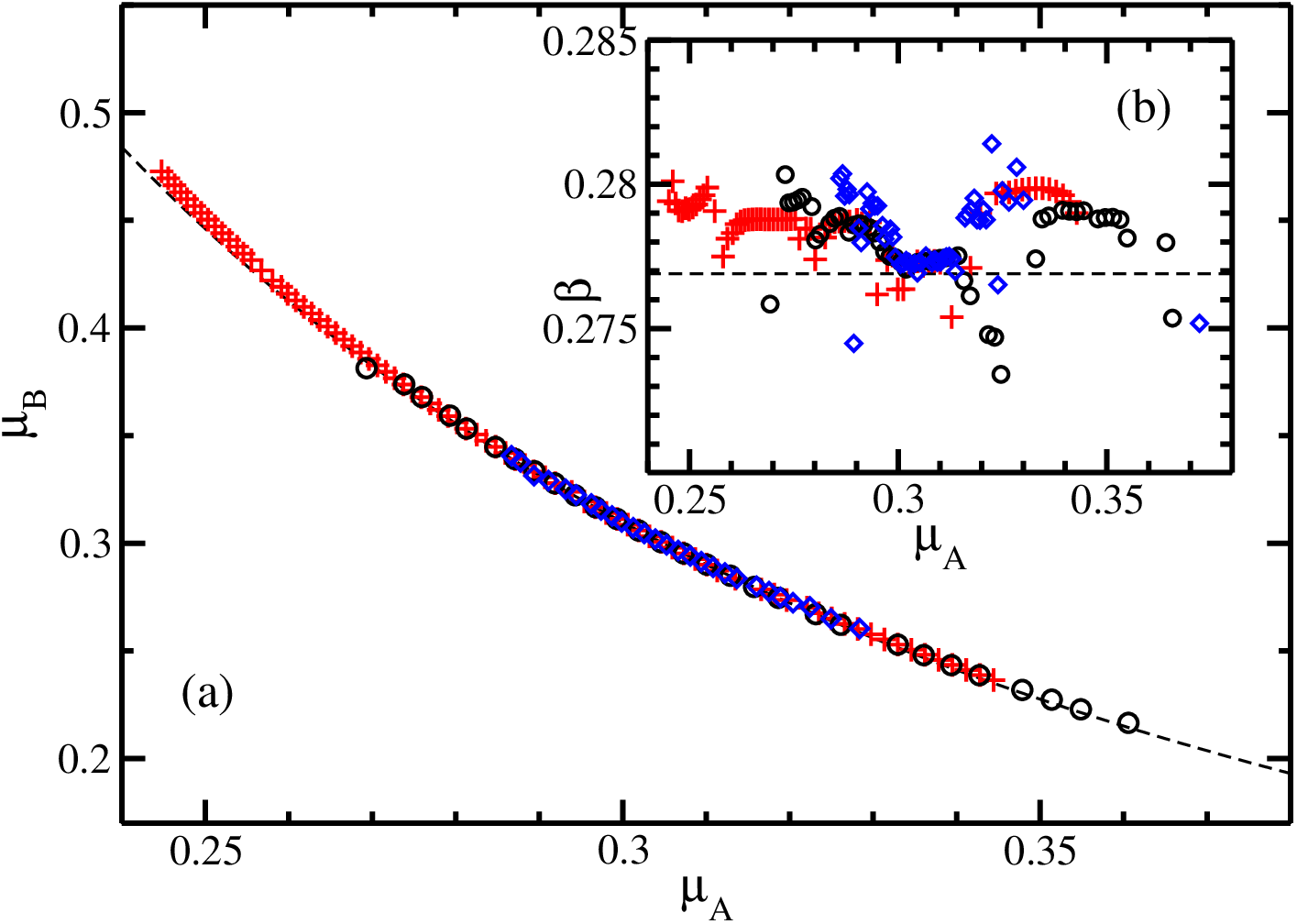}}
  \end{center}
  \caption{(Color online) Comparison of the critical values (a) and critical
    exponents $\beta$ (b) of the $AAB$ system obtained from
    series expansions up to order $N=24$ with different transformations, T1 ($\circ$), T2 ($+$) and
    T3 ($\diamond$).
    The dashed lines (\dashline) are given by Eq.~(\ref{eq:critline}) in (a)
    and by $\beta$ at the homogeneous critical
    point, $\beta = 0.2769$, from series expansions~\cite{jensen_93} for the
    CP, in (b). 
  } 
  \label{fig:comparison-AAB-T1+T2}
\end{figure}
%%%%%%%%%%%%%%%%%%%%%%%%%%%%%%%%%%%%%%%%%%%%%%%%%%%%%%%%%%%%%%%%%%%%%%%%%%%%%%%%%%%%%%%%%
%
%%%%%%%%%%%%%%%%%%%%%%%%%%%%%%%%%%%%%%%%%%%%%%%%%%%%%%%%%%%%%%%%%%%%%%%%%%%
\begin{figure}
  \begin{center}
    \scalebox{0.35}{\includegraphics[angle=0]{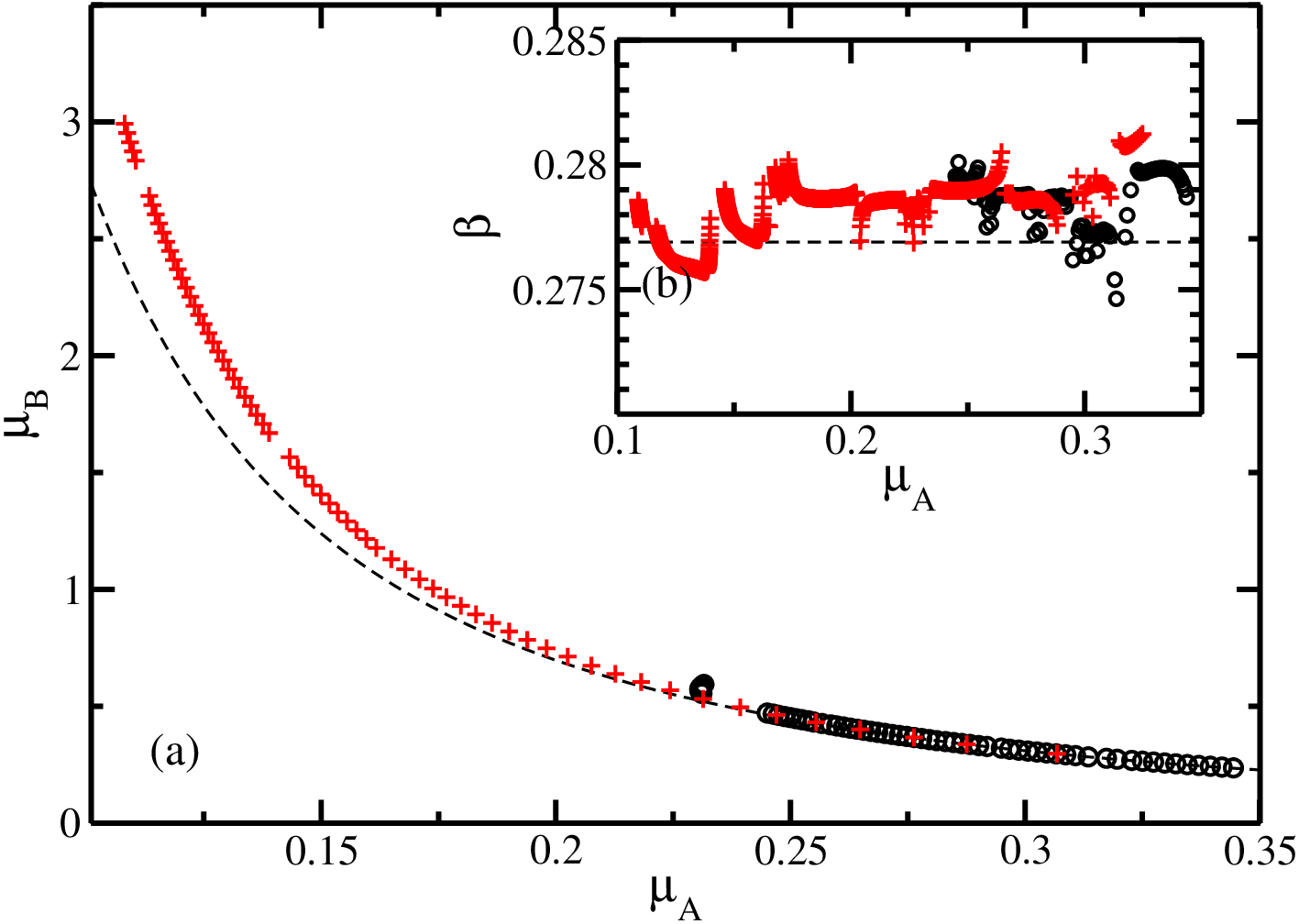}}
  \end{center}
  \caption{(Color online) Comparison of critical values (a) and critical exponents (b) of the
   $AAB$ lattice obtained from series expansions up to
   overall order $N = 24$ which are cut off at order $M = 12$
   ($+$) or in which all the terms up to order $M = 23$ ($\circ$) are
   retained.
   Transformation T2 was used to obtain these critical points.
   The dashed lines (\dashline) are given by Eq.~(\ref{eq:critline}) in (a)
    and by $\beta$ at the homogeneous critical
    point, $\beta = 0.2769$, from series expansions~\cite{jensen_93} for the
    CP, in (b).
  } 
  \label{fig:critline+beta-AAB-12+23}
\end{figure}
%%%%%%%%%%%%%%%%%%%%%%%%%%%%%%%%%%%%%%%%%%%%%%%%%%%%%%%%%%%%%%%%%%%%%%%%%%%%%%%%%%%%%%%%%
%
%%%%%%%%%%%%%%%%%%%%%%%%%%%%%%%%%%%%%%%%%%%%%%%%%%%%%%%%%%%%%%%%%%%%%%%%%%%
\begin{figure}
  \begin{center}
    \scalebox{0.35}{\includegraphics[angle=0]{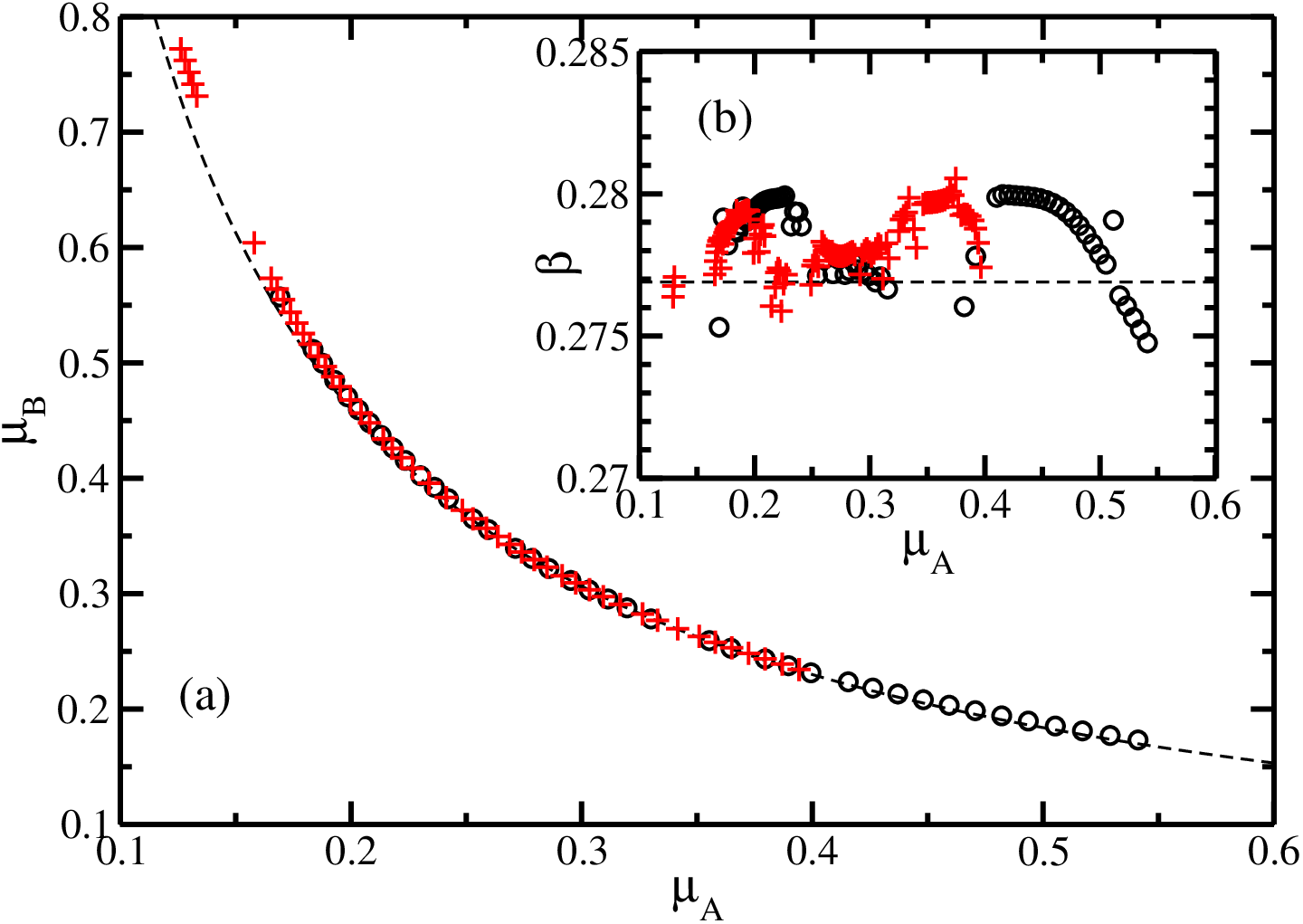}}
  \end{center}
  \caption{(Color online) Comparison of critical values (a) and critical exponents (b) of the 
   $AB$ lattice obtained from series expansions up to
   overall order $N = 24$ which are cut off at order $M = 12$
   ($+$) or in which all the terms up to order $M = 23$ ($\circ$) are
   retained.
   Transformations T2 and T1, respectively, were used to obtain these critical points.
   The dashed lines (\dashline) are given by Eq.~(\ref{eq:critline}) in (a)
    and by $\beta$ at the homogeneous critical
    point, $\beta = 0.2769$, from series expansions~\cite{jensen_93} for the
    CP, in (b).
  } 
 \label{fig:critline+beta-AB-12+23}
\end{figure}
%%%%%%%%%%%%%%%%%%%%%%%%%%%%%%%%%%%%%%%%%%%%%%%%%%%%%%%%%%%%%%%%%%%%%%%%%%%%%%%%%%%%%%%%%5
%
%%%%%%%%%%%%%%%%%%%%%%%%%%%%%%%%%%%%%%%%%%%%%%%%%%%%%%%%%%%%%%%%%%%%%%%%%%%%
\begin{figure}
  \begin{center}
    \scalebox{0.35}{\includegraphics[angle=0]{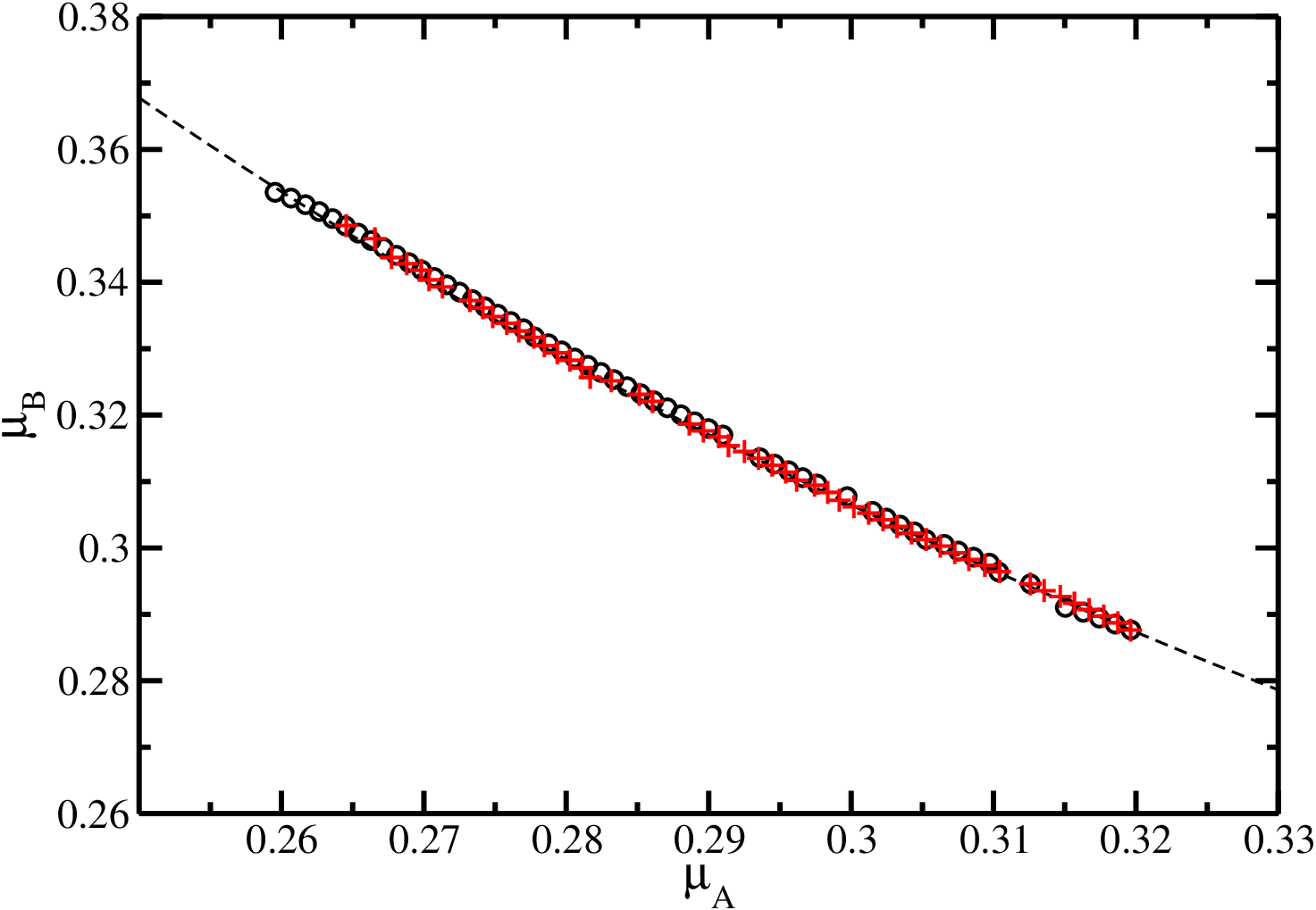}}
  \end{center}
  \caption{(Color online) Comparison of phase diagram for the $AB$ system obtained from series
    expansion up to order $N = 12$ ($\circ$) and for the $AABB$ chain obtained
    from series expansion up to order $N=24$ ($+$). 
    The dashed line (\dashline) is given by Eq.~(\ref{eq:critline}). 
  } 
  \label{fig:comparison-AB-12-AABB-24}
\end{figure}
%%%%%%%%%%%%%%%%%%%%%%%%%%%%%%%%%%%%%%%%%%%%%%%%%%%%%%%%%%%%%%%%%%%%%%%%%%%%%%%%%%%%%%%%%
%                      PDA
%
%%%%%%%%%%%%%%%%%%%%%%%%%%%%%%%%%%%%%%%%%%%%%%%%%%%%%%%%%%%%%%%%%%%%%%%%%%%
\begin{figure}
  \begin{center}
    \scalebox{0.35}{\includegraphics[angle=0]{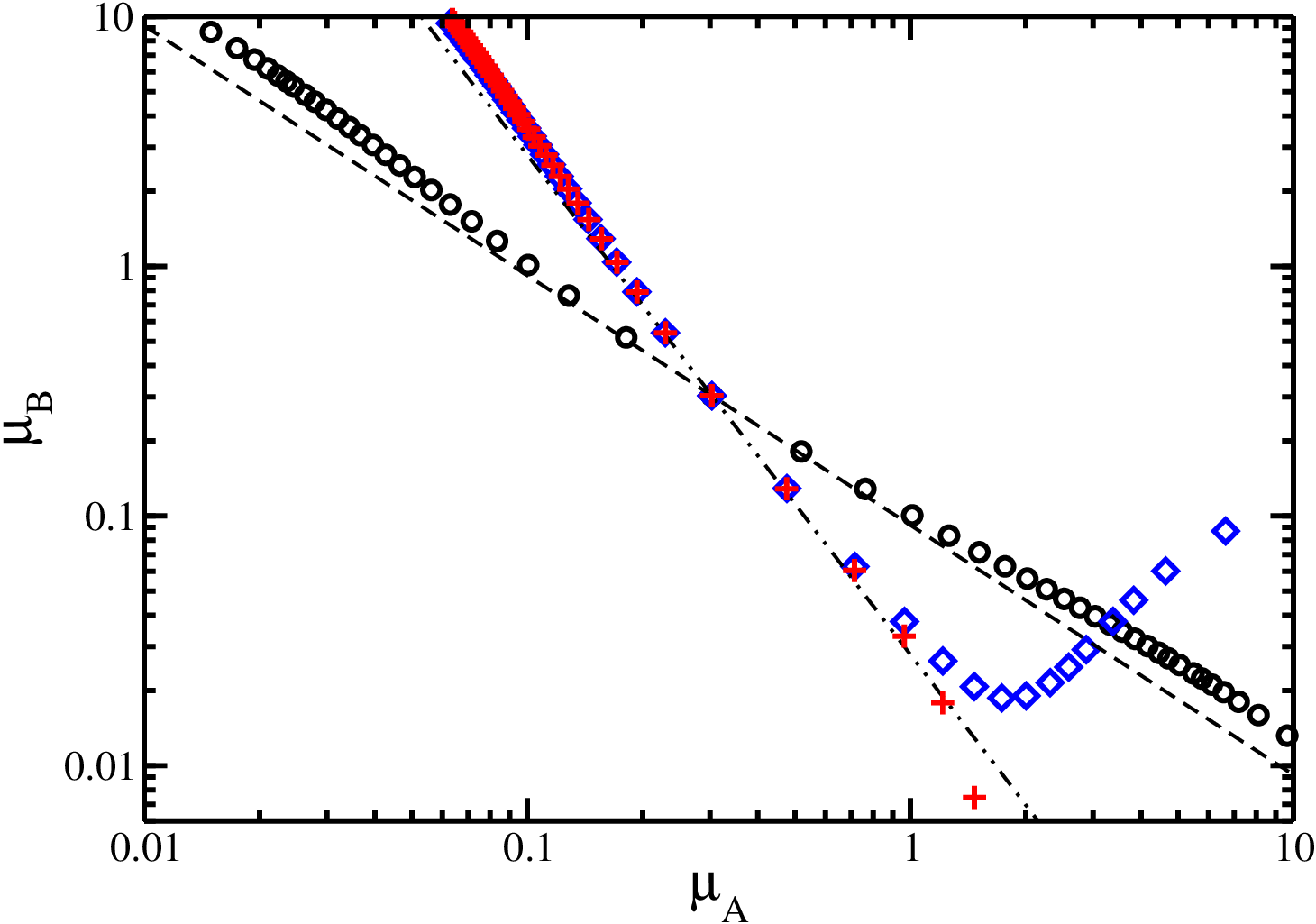}}
  \end{center}
  \caption{(Color online) Critical lines in $\log$-$\log$ scale for the $AB$ lattice ($\circ$) and
    the $AAB$ lattice ($+$ and $\diamond$) as obtained by PDA's: 
    The dashed lines (\dashline) and (\ddashline) are given by
    Eq.~(\ref{eq:critline}) for the $AB$- and the $AAB$-system, respectively.
  } 
  \label{fig:pda-AB-AAB}
\end{figure}
%%%%%%%%%%%%%%%%%%%%%%%%%%%%%%%%%%%%%%%%%%%%%%%%%%%%%%%%%%%%%%%%%%%%%%%%%%%%%%%%%%%%%%%%%

First, we analyzed periodic lattices, i.e.\ the CP in systems with a repeating
pattern of nodes characterized by the recovery rates $\mu_A$ and $\mu_B$. 
Thinking of a periodic system in terms of
unit cells that are repeated throughout the lattice and denoting a site in
this unit cell that has recovery rate $\mu_A$ ($\mu_B$) as an $A$($B$)-site,
the three $1d$-lattices $AB$, $AAB$ and $AABB$ are considered.
As can be easily seen from Eq.~(\ref{eq:critline}), the critical
lines of $AB$ and $AABB$ should coincide, at least sufficiently close to the
homogeneous critical point.
Therefore, we will only consider $AB$ and $AAB$ in detail, except for
a comparison of the stability of the critical values away from the homogeneous
critical point for $AB$ and $AABB$, which will be the subject of the last part
of this section.  

%%%%%%%%%%%%%%%%%%%%%%%%%%%%%%%%%%%%%%%%%%%%%%%%%%%%%%%%%%%%%%%%%%%%%%%%%%%%%%%%%%%%%%%%%%%%%%%%

\subsubsection{Results obtained by NPA's}

The series expansion for all three systems has been calculated up to order $N
= 24$. 
In order to evaluate how the estimates provided by the analysis described in
Sec.~\ref{sec:pade} behave with the order of the expansion, in
Figs.~\ref{fig:comparison-AB} (a) and (b) the critical
line and the critical exponent for the $AB$-lattice are shown for four
different values of $N$, $N = \{10, 17, 24\}$. 
The points in Figs.~\ref{fig:comparison-AB} (a) and (b)
were obtained after using transformation T1.
Most notably, as seen in Fig.~\ref{fig:comparison-AB}, the range of reliable
critical points increases with increasing order $N$.
Of course, the accuracy of the prediction
also increases with increasing $N$.
However, in general it can be said that even for low orders of the expansion
the critical values are of good accuracy.

The critical exponents are more sensitive to the order of expansion --
their behavior with $N$ is shown in the Fig.~\ref{fig:comparison-AB} (b).
Not surprisingly, we find that with increasing order the critical exponents
come closer to the best known value for the homogeneous CP from series
expansions, $\beta \simeq 0.2769$~\cite{jensen_93} -- at least close to the
homogeneous critical point.
Further away from this point, all the critical exponents for all orders
fluctuate between $\beta = 0.275$ and $\beta = 0.28$, the range in $\mu_A$ of reliable
exponents coinciding with the range for the critical points.

In an attempt to extend the range of applicability of the series expansion to
locate critical points, the linear transformations described in
Sec.~\ref{sec:pade} are applied to the expansion variables.
These transformations change the magnitude of the new variables $\bar{\mu}$
and $\delta$ and are therefore expected to provide extensions in different
regions of the phase diagram.
First, the transformations T1 and T2 are in employed in the $AB$ system (T3
being related by symmetry to T2). 
Fig.~\ref{fig:comparison-AB-T1+T2} shows
that for this particular lattice the transformation T1 performs far better
than T2 in 
giving stable critical points and critical exponents.
The results of the transformations for the asymmetric system $AAB$ are
compared in Figs.~\ref{fig:comparison-AAB-T1+T2} (a) and (b).
Here, two things are to note: transformation T2 extends the critical line
further in the $\mu_A < \mu_{\text{c}}$ direction than T1 and T3 does in
general poorly compared to the other two.

We also investigate the effects of different orders of Pad\'e approximants
for $\bar{\mu}$ and $\delta$ in the analysis as described in
Sec.~\ref{sec:pade}.
In order to do this, the steps given by Eqs.~(\ref{eq:step1}) and
(\ref{eq:step2}) are followed, but then, instead of taking the Pad\'e
approximant $\left[\floor{(N-1)/2}, \ceil{(N-1)/2} \right]_{\bar{\mu}}$ as
shown in Eq.~(\ref{eq:step3}), the series in Eq.~(\ref{eq:step2}) is truncated
at order $M < N-1$ in $\bar{\mu}$. 
This means that the coefficients of the terms $\bar{\mu}^n$ with $n \leq M$
that remain in the series are 
Pad\'e approximants in $\delta$ formed from polynomials of degree larger than
$N-1-M$.
Therefore, the extrapolation provided by the Pad\'e approximants of these
coefficients can be expected to be
more accurate than for polynomials of lower degree, which is the case for the
coefficients of $\bar{\mu}^n$ for $n > M$.
Then, the Pad\'e approximant $\left[\floor{M/2}, \ceil{M/2}
\right]_{\bar{\mu}}$ is taken.
The results of such a cut-off is that the range of the convergence of the
series is improved 
allowing for well behaved poles further away from the homogeneous critical
point.
The extension of the range for the phase-separation line achieved by this
procedure can be seen in Figs.~\ref{fig:critline+beta-AAB-12+23} (a) and
\ref{fig:critline+beta-AAB-12+23} (b) for the $AAB$-system.
For this system, the transformation T2 significantly extends the
line of critical points into the region
$\mu_A^{(\text{c})} < \mu_{\text{c}}$ and $\mu_B^{(\text{c})} >
\mu_{\text{c}}$. 
In Fig.~\ref{fig:critline+beta-AAB-12+23} (b), it can be seen that the critical
exponents deviate
from the value $\beta = 0.2769$ by less than $1.2 \%$.

The fact that this procedure of truncating in orders of expansion in
$\bar{\mu}$ performs well in this particular case can be understood in the
following manner.
For the transformation T2, $\bar{\mu} = \mu_A$ and $\delta = \mu_B - \bar{\mu}
= \mu_B - \mu_A$, leading to small values of the variable ($\bar{\mu}$) in
which the expansion is cut, while $\delta$ is very large for $\mu_A <
\mu_{\text{c}}$ and $\mu_B > \mu_{\text{c}}$.
Therefore, we can expect the effects of dropping terms in $\bar{\mu}$ to be
small while the benefit of only using Pad\'e approximants in $\delta$ formed
from high-order polynomials to be large.
This explains the sizeable extension of the region in which the poles of the
series are well-behaved.

In contrast to the $AAB$ system, the procedure described above does not
produce much extension of the critical line for the $AB$ lattice, as can be
seen Fig.~\ref{fig:critline+beta-AB-12+23}: there are only a few points
obtained by using transformation T2 further away from the homogeneous point
than the critical points by the regular analysis with T1.

It is worth noting that the deviation of critical line from the line given by
Eq.~(\ref{eq:critline}) appreciably increases away from the homogeneous
critical point (see Fig.~\ref{fig:critline+beta-AAB-12+23}).
It shows that Eq.~(\ref{eq:critline}) is only a lowest-order approximation in
relation to the homogeneous critical point.

For the $AABB$ system, it is found that, in general, there is a smaller range
of the critical values and critical exponents compared to the situation for
$AB$ and $AAB$ chains.
In order to investigate this difference in range of stability of the critical
points and 
exponents, the $AABB$ critical line is compared with that for the
$AB$ lattice, which should follow the same curve according to
Eq.~(\ref{eq:critline}).
It is found that the difference of the stability of the critical values seems
to be a result of the  
number of unit cells that the series expansion takes into account.
For order $N = 24$, six unit cells on either side of the origin affect the
series expansion for $AABB$ while the number is twice that for $AB$.
If order $N=12$ expansion for the $AB$-lattice, which also only takes six unit
cells into account, 
is compared with the order $N=24$ expansion for the $AABB$ chain, then we find
that their ranges are very similar. 
This is shown in Fig.~\ref{fig:comparison-AB-12-AABB-24} where the critical
points start to 
fluctuate wildly for both system at the same point in the phase diagrams 
(not presented in Fig.~\ref{fig:comparison-AB-12-AABB-24}).

%%%%%%%%%%%%%%%%%%%%%%%%%%%%%%%%%%%%%%%%%%%%%%%%%%%%%%%%%%%%%%%%%%%%%%%%%%%%%%%%%%%%%%

\subsubsection{Results obtained by PDA's \label{pda:hetero}}

As described in Sec.~\ref{sec:pda}, by applying the PDA method to the
$P_\infty (\mu_A, \mu_A)$ we can also compute a line of critical
points.
The results obtained by this method depend on the choices of the label sets, 
$\mathbf{L}$, $\mathbf{M}$, $\mathbf{N}$ for the polynomials
$P_{\mathbf{L}}(x,y)$, $Q_{\mathbf{M}}(x,y)$ and $R_{\mathbf{N}}(x,y)$
and the matching set $\mathbf{K}$.
For all our analyses employing PDA's, we use either label sets
$\mathbf{M}$, $\mathbf{N}$ and $\mathbf{K}$ that have
triangular or rectangular form.
The label set $\mathbf{L}$ is then chosen to mimic the form of the
others while at the same time making sure that the number of elements
in all four sets satisfies the constraint that is imposed on them (see
Sec.~\ref{sec:pda}).
By ''triangular'' label set $\mathbf{M}$ we mean that $(i,j) \in \mathbf{M}$
if $0 \leq (i + j)
\leq M_{max}$ while ''rectangular'' refers to a set $\mathbf{M}$ for
which $(i,j) \in \mathbf{M}$ if $0 \leq i \leq M_{max,i}$ and 
$0 \leq j \leq M_{max,j}$ with some integers $M_{max}$, $M_{max,i}$
and $M_{max,j}$.
From here on, we will refer to a particular choice of the label sets
$\mathbf{L}$, $\mathbf{M}$, $\mathbf{N}$ and matching set
$\mathbf{K}$ as an \emph{input set}.

Often, different input sets produce critical lines of varying
extent in $(\mu_A, \mu_B)$-space.
%However, they usually agree on the loci of critical points.
In Fig.~\ref{fig:pda-AB-AAB}, we show the results from one input set
for the $AB$ lattice and from two input sets for the $AAB$ lattice in
comparison to the lines given by Eq.~(\ref{eq:critline}).
The input set for the $AB$ lattice are triangular while the ones for
the $AAB$ system are rectangular and triangular.
The critical lines obtained by PDA's are presented in $\log$-$\log$ scale because 
they extend to a rather wider range than
those calculated by NPA's (cf.\
Figs.~\ref{fig:comparison-AB}, \ref{fig:comparison-AAB-T1+T2} and \ref{fig:pda-AB-AAB}).
One can clearly see that around the homogeneous critical point the
lines from the series and described by Eq.~(\ref{eq:critline}) agree very well
while deviations develop further away.
It can be seen in Fig.~\ref{fig:pda-AB-AAB} that the critical lines for
triangular and rectangular input sets in the case of $AAB$ chain coincide in
the vicinity of the critical point with a consistent tendency to be above the
prediction given by Eq.~(\ref{eq:critline}).
There is a point though where these two approximate curves for the critical line diverge:
from there, the estimates given by the PDA's can no longer be considered reliable
(cf.\ the behavior of the curves marked by $+$ and $\diamond$ in
Fig.~\ref{fig:pda-AB-AAB} for $\mu_A \agt 1$).

\subsection{Disordered Contact Process with Bimodal Distribution}

\subsubsection{Results obtained by NPA's}

%
%%%%%%%%%%%%%%%%%%%%%%%%%%%%%%%%%%%%%%%%%%%%%%%%%%%%%%%%%%%%%%%%%%%%%%%%%%%
\begin{figure}
  \begin{center}
    \scalebox{0.35}{\includegraphics[angle=0]{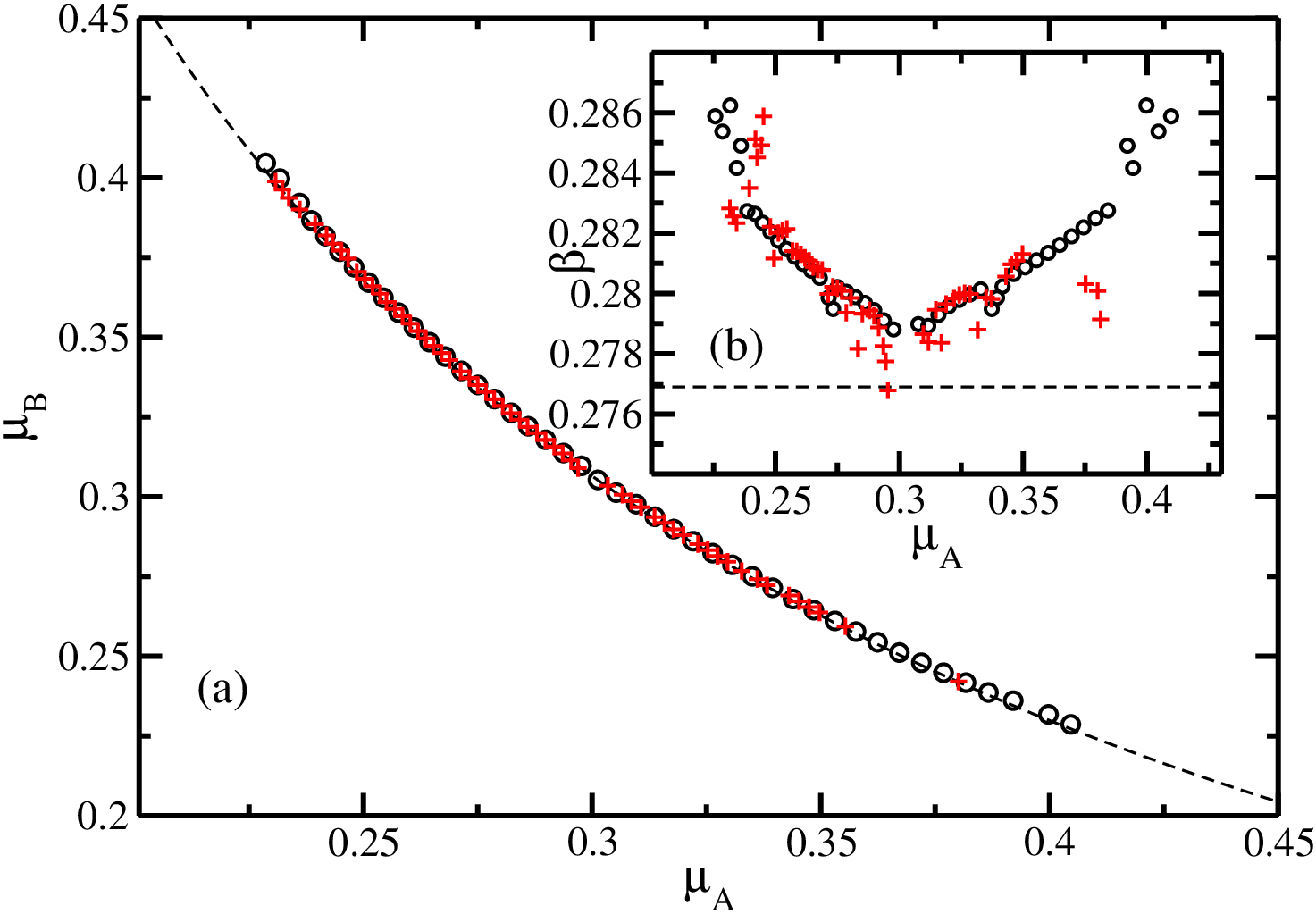}}
  \end{center}
  \caption{(Color online) Comparison of critical values (a) and critical exponents (b) of the disordered system with
   $p = 0.5$ obtained from series expansions up to order $N = 12$ using
   transformations T1 ($\circ$) and T2 ($+$).
   The dashed lines (\dashline) denote (a) the curve given by Eq.~(\ref{eq:critline}) and
   (b) the value for the critical exponent at the homogeneous critical point for the CP from series
   expansion, $\beta=0.2769$ \cite{jensen_93}.
 } 
 \label{fig:critline+beta-p0.5-11-12-T1+T2}
\end{figure}
%%%%%%%%%%%%%%%%%%%%%%%%%%%%%%%%%%%%%%%%%%%%%%%%%%%%%%%%%%%%%%%%%%%%%%%%%%%%%%%%%%%%%%%%%5
%
%%%%%%%%%%%%%%%%%%%%%%%%%%%%%%%%%%%%%%%%%%%%%%%%%%%%%%%%%%%%%%%%%%%%%%%%%%%
\begin{figure}
  \begin{center}
    \scalebox{0.35}{\includegraphics[angle=0]{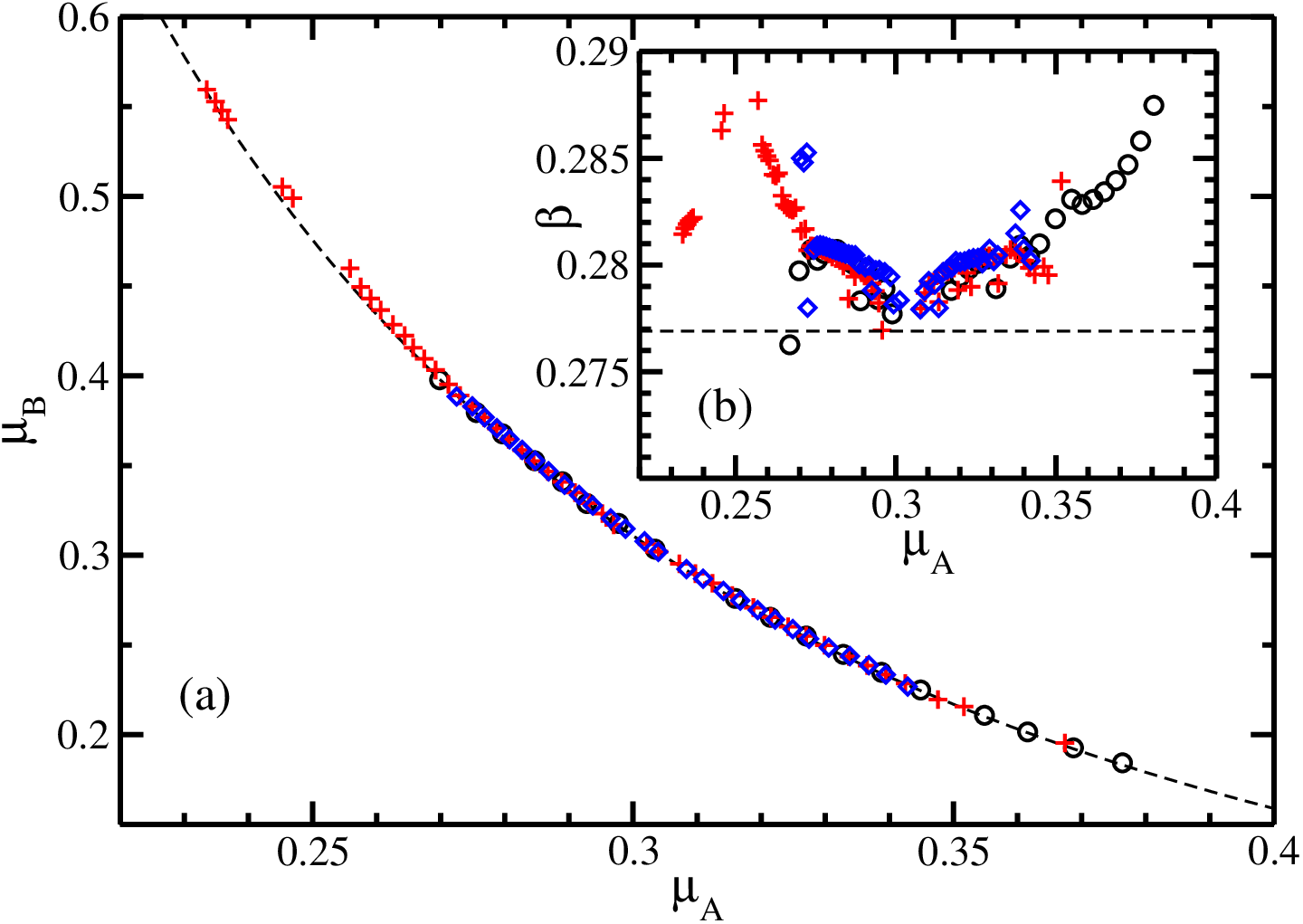}}
  \end{center}
  \caption{(Color online) Comparison of critical values (a) and critical exponents (b) of the disordered
   system with $p=0.7$ obtained from series expansions up to
   order $N = 12$: T1 ($\circ$), T2 ($+$) and T3 ($\Diamond$).
   The dashed lines (\dashline) denote (a) the curve given by
   Eq.~(\ref{eq:critline}) and
   (b) the value for the critical exponent at the homogeneous critical point for the CP from series
   expansion, $\beta=0.2769$ \cite{jensen_93}.
 } 
 \label{fig:critline+beta-p0.3-11-12-T1+T2}
\end{figure}
%%%%%%%%%%%%%%%%%%%%%%%%%%%%%%%%%%%%%%%%%%%%%%%%%%%%%%%%%%%%%%%%%%%%%%%%%%%%%%%%%%%%%%%%%
%
%%%%%%%%%%%%%%%%%%%%%%%%%%%%%%%%%%%%%%%%%%%%%%%%%%%%%%%%%%%%%%%%%%%%%%%%%%%
\begin{figure}
  \begin{center}
    \scalebox{0.35}{\includegraphics[angle=0]{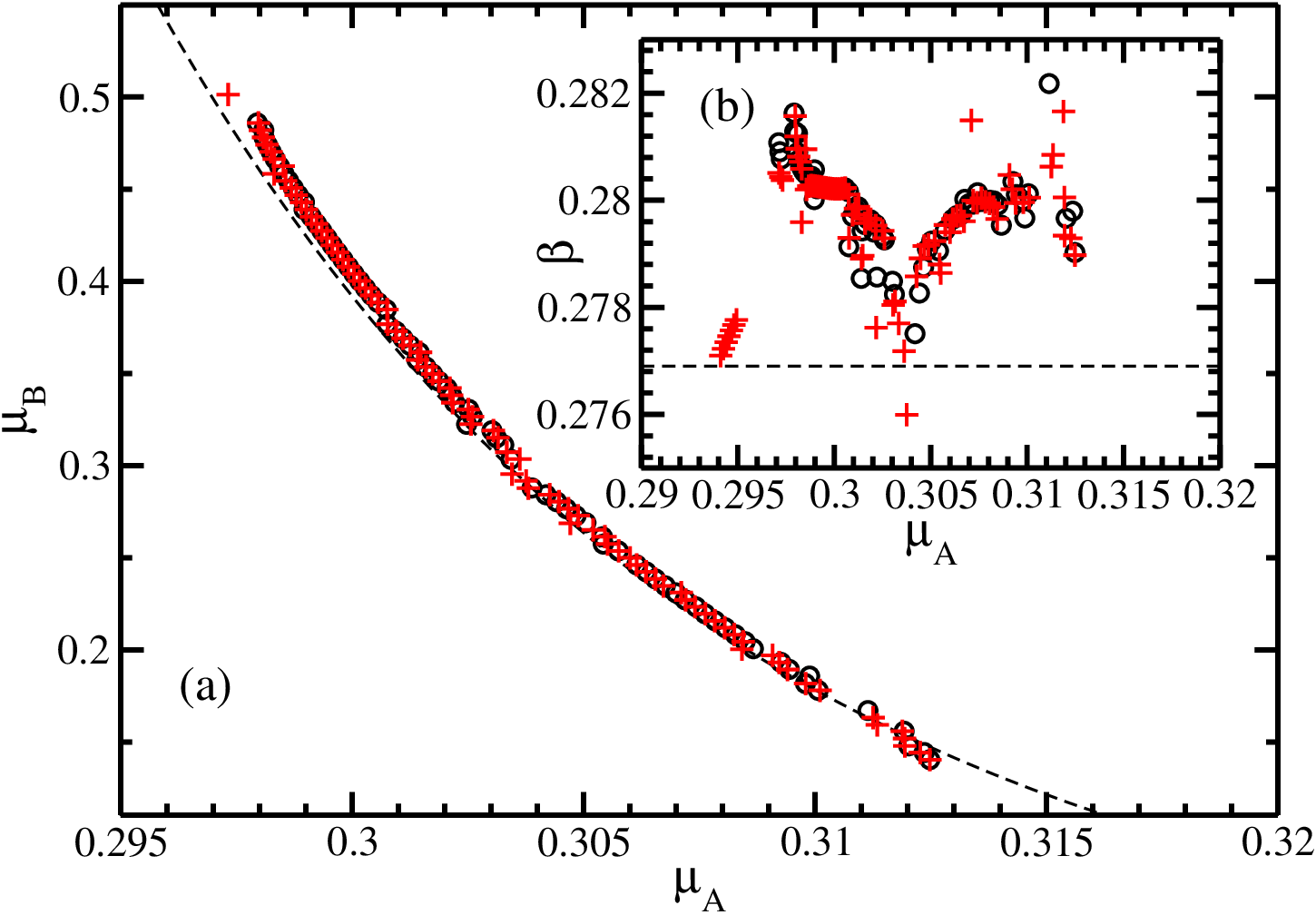}}
  \end{center}
  \caption{(Color online) Comparison of critical values (a) and
    critical exponents (b) of the disordered $1d$ lattice with $p=0.96$
   obtained from series expansions up to order $N = 12$ using exact configurational averaging
   ($\circ$) and approximate configurational averaging
   ($+$) with $k_{\text{max}}(12,p) = 2$.
   In both expansions T1 was used.
   The dashed lines (\dashline) are given by Eq.~(\ref{eq:critline}) in (a)
   and by $\beta$ at the homogeneous critical
   point, $\beta = 0.2769$, from series expansions~\cite{jensen_93} for the
   CP, in (b).
  } 
  \label{fig:12-kmax2}
\end{figure}
%%%%%%%%%%%%%%%%%%%%%%%%%%%%%%%%%%%%%%%%%%%%%%%%%%%%%%%%%%%%%%%%%%%%%%%%%%%%%%%%%%%%%%%%%
%
%%%%%%%%%%%%%%%%%%%%%%%%%%%%%%%%%%%%%%%%%%%%%%%%%%%%%%%%%%%%%%%%%%%%%%%%%%%
\begin{figure}
  \begin{center}
    \scalebox{0.35}{\includegraphics[angle=0]{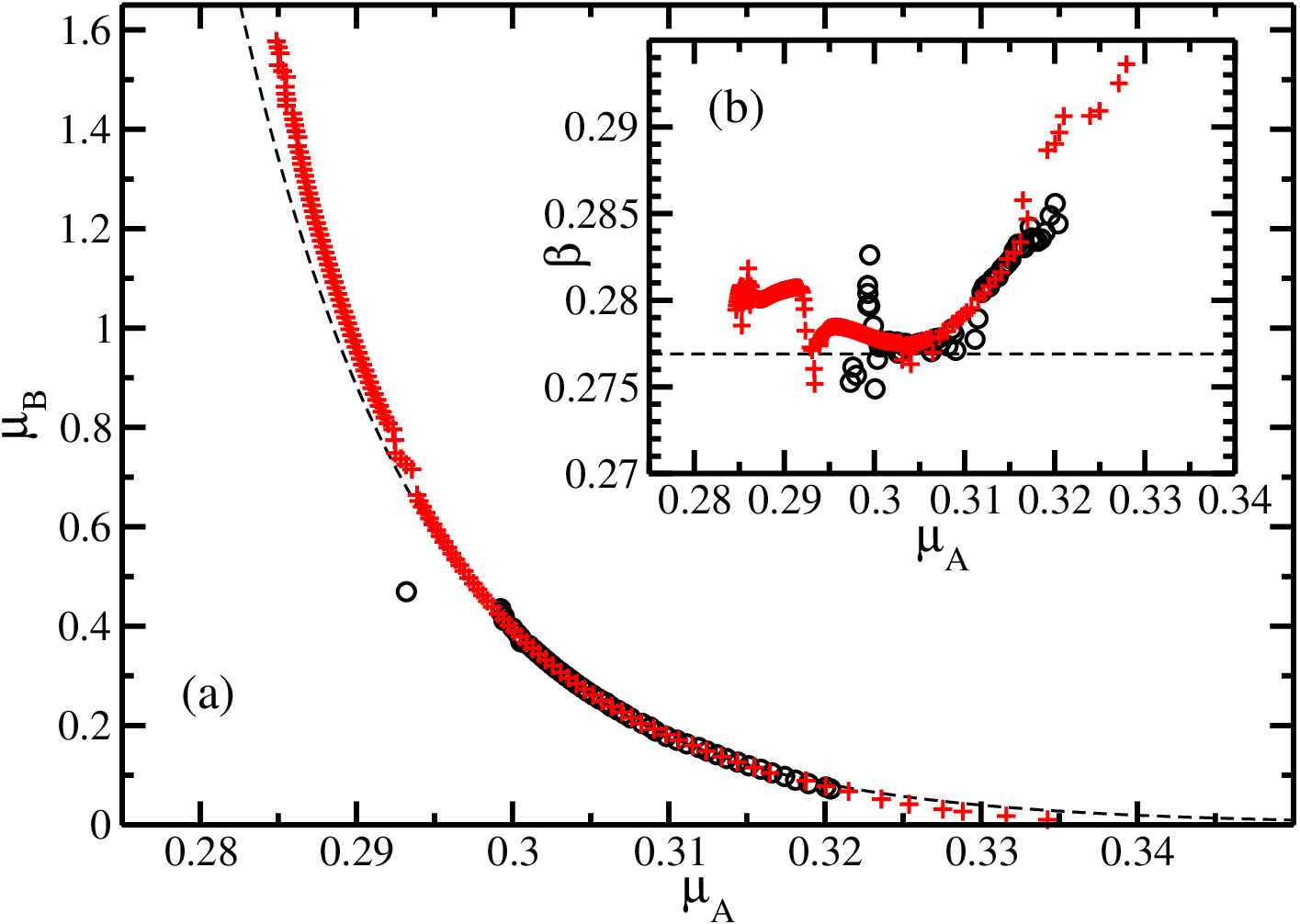}}
  \end{center}
  \caption{(Color online) Comparison of critical values (a) and critical exponents (b) of the
   disordered $1d$ lattice with $p=0.96$ obtained from series expansions of the
   approximately configurationally averaged survival probability up to
   order $N = 19$ using the transformations T1 ($\circ$) and T2 ($+$).
   The dashed lines (\dashline) denote (a) the curve given by Eq.~(\ref{eq:critline}) and
   (b) the value for the critical exponent at the homogeneous critical point for the CP from series
   expansion, $\beta=0.2769$~\cite{jensen_93}.
 } 
 \label{fig:critline+beta-p0.96-18-19-T1+T2}
\end{figure}
%%%%%%%%%%%%%%%%%%%%%%%%%%%%%%%%%%%%%%%%%%%%%%%%%%%%%%%%%%%%%%%%%%%%%%%%%%%%%%%%%%%%%%%%%5
%%
%%%%%%%%%%%%%%%%%%%%%%%%%%%%%%%%%%%%%%%%%%%%%%%%%%%%%%%%%%%%%%%%%%%%%%%%%%%
\begin{figure}
  \begin{center}
    \scalebox{0.35}{\includegraphics[angle=0]
      {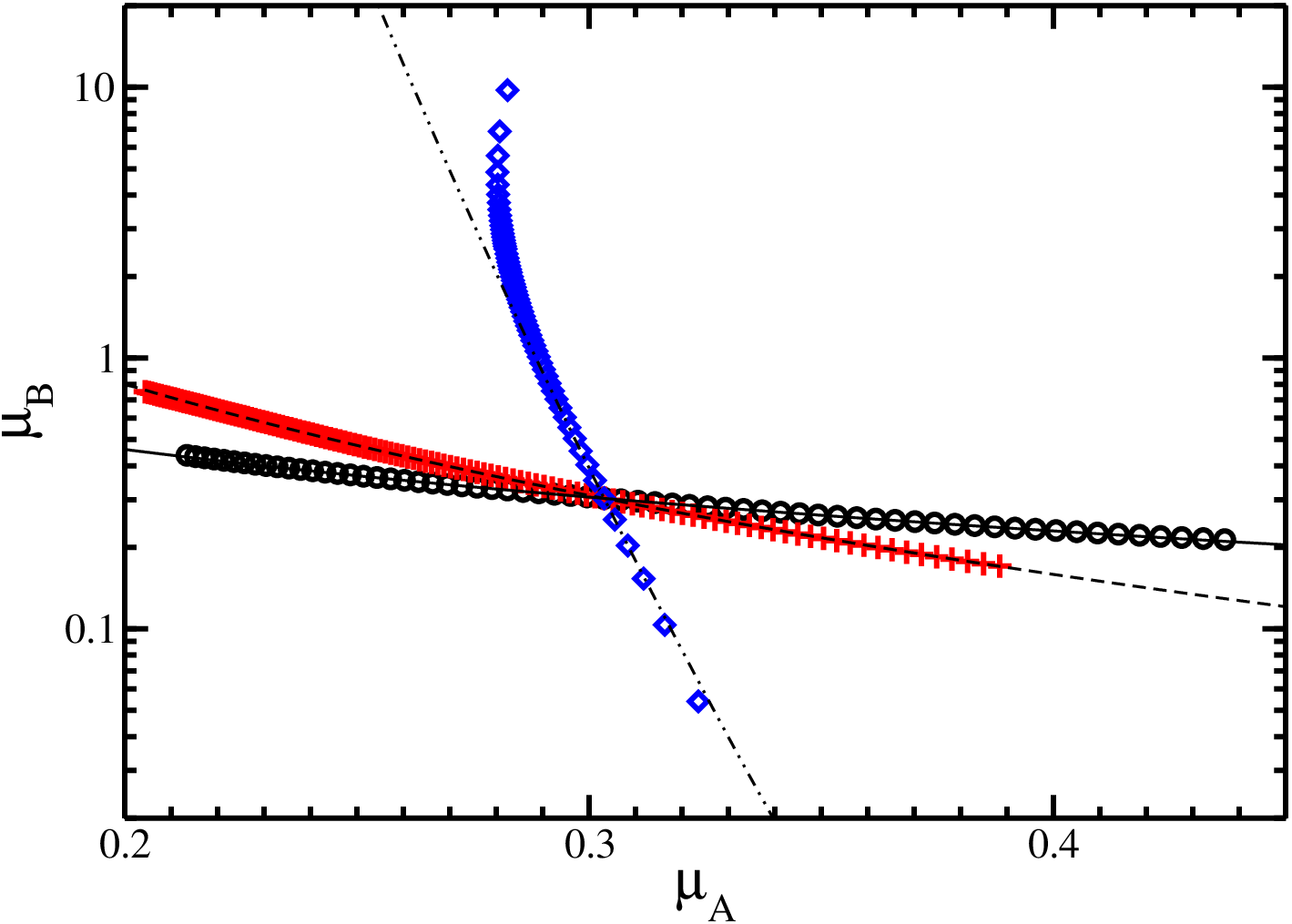}}
  \end{center}
  \caption{(Color online) Critical lines in $\log$-linear scale for the
    disordered lattice with $p = 0.5$ ($\circ$), $p=0.7$ ($+$) and
    $p=0.96$ ($\diamond$) as obtained by PDA's: 
    The three lines (\sline), (\dashline) and (\ddashline) are given by
    Eq.~(\ref{eq:critline}) for the three disordered systems characterized by
    the values $p=0.5$, $p=0.7$ and $p=0.96$, respectively.
  } 
 \label{fig:pda-critline-p0.5+p0.3+p0.04}
\end{figure}
%%%%%%%%%%%%%%%%%%%%%%%%%%%%%%%%%%%%%%%%%%%%%%%%%%%%%%%%%%%%%%%%%%%%%%%%%%%%%%%%%%%%%%%%%5
%
%
%
For the disordered CP, in which the recovery rates are drawn from the bimodal
distribution given by Eq.~(\ref{eq:pdf}), the same analysis as
for the periodic lattices is carried out, with the only difference that the
survival probability is configurationally averaged. 
Up to order $N=12$, the expansion of the survival probability
can be configurationally averaged exactly for any value of $p$.
The results for such systems characterized by $p=0.5$ and
$p=0.7$ are shown in Figs.~\ref{fig:critline+beta-p0.5-11-12-T1+T2} and
\ref{fig:critline+beta-p0.3-11-12-T1+T2}. 
For both values of $p$, the line of critical points and the critical
exponents for the the two different linear transformations, T1 and T2, are
compared.

Similar to the heterogeneous $AB$ lattice, we find that for the disordered
system with $p=0.5$, transformation T1 extends the line of critical points
furthest (see Fig.~\ref{fig:critline+beta-p0.5-11-12-T1+T2}). 
The critical points are again well described by Eq.~(\ref{eq:critline}).
The critical exponents, however, behave differently for the disordered case
than for the heterogeneous lattices: while in the latter case they
fluctuate 
up immediately away from homogeneous point $\mu_{\text{c}}$ but then
fluctuate down again, in the former system, they almost linearly increase away
from the homogeneous critical point.

Fig.~\ref{fig:critline+beta-p0.3-11-12-T1+T2} shows the results for the
disordered lattice with $p=0.7$. 
This system behaves more like the heterogeneous $AAB$ lattice with
transformation T2 extending the critical line furthest to the left, for $\mu_A
< \mu_{\text{c}}$ and T1
furthest to the right, $\mu_A > \mu_{\text{c}}$.
As usual, the points obtained from T3 lie in the middle and thus do not extend
the critical line in any direction.
The critical exponent $\beta$ displays very similar behavior as for $p=0.5$,
mostly monotonically increasing away from the homogeneous critical point.

As noted in Sec.~\ref{sec:configav}, the exact configurational averaging of the
survival probability, $\langle \op \rangle$, becomes rapidly computationally
very demanding with order of expansion and is only possible for $N \leq 12$.
Therefore, the approximate scheme for averaging has to be applied
for orders $N > 12$.
Such a scheme has been described in Sec.~\ref{sec:configav}. 
In order to test the reliability of this scheme, we analyze the disordered
systems characterized by $p=0.96$, for which both exact and approximate
averaging, with $k_{\text{max}}(12,p) = 2$, are possible.
The comparison between the two configurational averages are shown in
Fig.~\ref{fig:12-kmax2} (a) and (b).
It can be seen that the critical points agree very
well for the two configurational averages where they both produce stable
critical points, showing that the approximation scheme is indeed faithful for
small disorder concentrations. 

For a system with small concentration of $B$ sites, e.g.\ for $p \geq 0.96$,
the approximate scheme described in Sec.~\ref{sec:configav} can be applied for
configurational averaging up to order $N=19$.
Below, we present results for $p=0.96$ with $k_{\text{max}}(19,p) = 2$.
Figs.~\ref{fig:critline+beta-p0.96-18-19-T1+T2} (a) and (b) show the critical
values and critical exponents for this case.
Clearly, the transformation T2 works best for this system as is to be expected
from its performance for the heterogeneous periodic $AAB$ lattice.
The critical points from transformation T3 cover less range than T1 or T2, so
we left these points out of Fig.~\ref{fig:critline+beta-p0.96-18-19-T1+T2}.
Deviation of the critical line from the curve given by Eq.~(\ref{eq:critline})
can be seen for 
$\mu_A < \mu_{\text{c}}$. 
The critical exponent $\beta$ increases monotonically with increasing value of
$\mu_A$ away from the homogeneous
critical point, $\mu_A > \mu_{\text{c}}$, while for $\mu_A < \mu_{\text{c}}$,
significant fluctuations can be seen.

It should be mentioned that the critical line is only marginally extended by
the procedure of cutting the 
series in different orders for $\bar{\mu}$ and $\delta$ described in
the previous section.

\subsubsection{Results obtained PDA's\label{pda:disorder}}

Using similar input sets as for the heterogeneous systems (as
described in Sec.~\ref{pda:hetero}), we apply PDA's to the disordered
systems as well. 
Generally, smaller input sets, with fewer elements in the label sets, are
used because the series are shorter for the disordered systems but the
triangular and rectangular shapes are still maintained.
In Fig.~\ref{fig:pda-critline-p0.5+p0.3+p0.04}, phase separation lines
for three systems with the different degree of disorder 
characterized by the following values of  $p = 0.5$,
$p=0.3$ and $p=0.96$ are shown. 
It can be clearly seen that Eq.~(\ref{eq:critline}) describes the
critical lines obtained by the PDA's very well around the homogeneous
point. 
For the $p=0.96$ system, from $\mu_B = 1.5$ the critical points deviate from
Eq.~(\ref{eq:critline}).
Consistent with the results from the NPA's, the
critical points obtained by PDA's deviate above the line given by
Eq.~(\ref{eq:critline}).

%%%%%%%%%%%%%%%%%%%%%%%%%%%%%%%%%%%%%%%%%%%%%%%%%%%%%%%%%%%%%%%%%%%%

\section{Discussion \& Conclusion \label{sec:DiscConc}}

To conclude, we have presented a detailed description of the supercritical
series expansions for the survival probability of the contact process in
heterogeneous and disordered one-dimensional binary lattices. 
The heterogeneous systems are modeled by 
lattices with repeating patterns of sites of two types $A$ and $B$ 
characterized by 
different recovery rates, $\mu_A$ and $\mu_B$, and the
disordered systems are represented by lattices of similar sites randomly
placed on the lattice sites with probabilities $p$ and $1-p$, for nodes $A$
and $B$, respectively.
For the analysis of the two-variable series (in $\mu_A$ and $\mu_B$), we have
presented a scheme based on NPA's in 
order to extract critical values and the critical exponent $\beta$
%.
and have also used PDA's to obtain
estimates for the line of critical points.

It has been demonstrated that 
(i) using symmetric and asymmetric linear
transformations, it is possible to extend the range of stable critical points
in different regions of the rate-space $(\mu_A, \mu_B)$;
(ii) keeping different orders in the Pad\'e approximants with respect to the
two variables, an extension of the line of critical points and an extended
range for the critical exponents can be achieved;
(iii) results from NPA's and PDA's compare well, PDA's usually widening the
range of the critical lines 
a bit further in the $(\mu_A, \mu_B)$-plane;
(iv) an approximate scheme for configurational averaging can be applied in
disordered lattices.

In general, the critical values can be reliably obtained by supercritical
series expansions and they are in good agreement with 
the analytical approximation given by Eq.~(\ref{eq:critline}).
For the critical exponents, the results are less conclusive due to larger
errors, but certainly give some indication as to what happens to the universal
critical behavior when spatial heterogeneity and disorder is introduced.
For all heterogeneous lattices, we see fluctuations away from the best known
value for the CP from series expansions, $\beta = 0.2769$~\cite{jensen_93},
but they hardly ever exceed $\beta = 0.28$ over the range of reliable points.
This suggests that the CP for heterogeneous lattices still belongs to
the DP universality class.

For the disordered systems, we see a qualitatively different picture.
In general, we find that the critical exponents monotonically increase away
from the value at the homogeneous critical point.
As we were only able to compute the exact configurationally averaged survival
probability up to order $N=12$ and an approximate one up to order $N=19$, we
do not have very high precision, but the above tendency is clearly visible in
all the data.  
This picture of continuously changing exponents is consistent with what a
number of other authors~\cite{moreira_96, cafiero_98, dickman_98,
  hooyberghs_03, hooyberghs_04} have found
and with the Harris criterion~\cite{harris_74_2}.

\section{Acknowledgments}

The authors would like to thank I.\ Jensen, J.\ Stilck and S.~V.~Fallert for their helpful
correspondence and remarks.
The computations were
partly performed on the Cambridge-Cranfield High-Performance Computer
facility (HPCF) and the Cambridge University Condor Project. 
CJN would like to thank the UK EPSRC and the Cambridge European Trust for financial
support.

\appendix

\section{Selected series coefficients}

%%%%%%%%%%%%%%%%%%%%%%%%%%%%%%%%%%%%%%%%%%%%%%%%%

\begin{table*}
  \caption{
    Series for the ultimate survival probability, $P_{\infty} = 1 -
    \sum_{n=1}^{N} \sum_{m=0}^n c_{nm} \mu_A^{n-m} \mu_B^m$, for the
    heterogeneous lattice $AB$ starting from a single occupied
    $A$-site up to order $N=15$ (continued in Tab.~\ref{table-AB-2}). 
    \label{table-AB-1}
  }
  \begin{ruledtabular}
    \begin{tabular}{ccccccccc}
      $n$ & $m$ & $c_{nm} $ & $n$ & $m$ & $c_{nm}$ &  $n$ & $m$ & $c_{nm} $ \\
      \hline
      0 &   0 & 1.00000000000000000000e+00    &    &   7 & 1.65000000000000000000e+01         &    &  12 & -0.00000000000000000000e+00        \\
        &     &                               &    &   8 & -0.00000000000000000000e+00        &    &     &                                    \\
      1 &   0 & -1.00000000000000000000e+00   &    &     &                                    & 13 &   0 & -1.00000000000000000000e+00        \\
        &   1 & -0.00000000000000000000e+00   &  9 &   0 & -1.00000000000000000000e+00        &    &   1 & -2.07000000000000000000e+02        \\
        &     &                               &    &   1 & -6.80000000000000000000e+01        &    &   2 & -6.74625000000000000000e+03        \\
      2 &   0 & 1.00000000000000000000e+00    &    &   2 & -7.58500000000000000000e+02        &    &   3 & -7.44351406860351562500e+04        \\
        &   1 & -2.00000000000000000000e+00   &    &   3 & -1.69084472656250000000e+03        &    &   4 & -3.01900451407580578234e+05        \\
        &   2 & -0.00000000000000000000e+00   &    &   4 & 3.02032038031683987356e+03         &    &   5 & -2.76109780525427486282e+05        \\
        &     &                               &    &   5 & 1.11290729437934101043e+03         &    &   6 & 8.00284240250960225239e+05         \\
      3 &   0 & -1.00000000000000000000e+00   &    &   6 & -1.77129394531249909051e+03        &    &   7 & 4.74728979742886382155e+05         \\
        &   1 & 5.00000000000000000000e-01    &    &   7 & -4.27500000000000000000e+02        &    &   8 & -3.91561916647056990769e+05        \\
        &   2 & -1.50000000000000000000e+00   &    &   8 & -2.10000000000000000000e+01        &    &   9 & -2.26055896148412110051e+05        \\
        &   3 & -0.00000000000000000000e+00   &    &   9 & -0.00000000000000000000e+00        &    &  10 & -3.99103976440429469221e+04        \\
        &     &                               &    &     &                                    &    &  11 & -2.51875000000000000000e+03        \\
      4 &   0 & 1.00000000000000000000e+00    & 10 &   0 & 1.00000000000000000000e+00         &    &  12 & -4.40000000000000000000e+01        \\
        &   1 & 3.00000000000000000000e+00    &    &   1 & 9.40000000000000000000e+01         &    &  13 & -0.00000000000000000000e+00        \\
        &   2 & -1.20000000000000000000e+01   &    &   2 & 1.45700000000000000000e+03         &    &     &                                    \\
        &   3 & 3.50000000000000000000e+00    &    &   3 & 6.01312646484374818101e+03         & 14 &   0 & 1.00000000000000000000e+00         \\
        &   4 & 1.02500000000000000000e+01    &    &   4 & -7.55620689863017105381e+01        &    &   1 & 2.58000000000000000000e+02         \\
        &     &                               &    &   5 & -1.94405384114583393966e+04        &    &   2 & 1.02715000000000000000e+04         \\
      5 &   0 & -1.00000000000000000000e+00   &    &   6 & 4.71605225664304089150e+03         &    &   3 & 1.41998072448730497854e+05         \\
        &   1 & -9.00000000000000000000e+00   &    &   7 & 4.74978173828125090949e+03         &    &   4 & 7.83967321483503794298e+05         \\
        &   2 & 1.02500000000000000000e+01    &    &   8 & 7.19500000000000000000e+02         &    &   5 & 1.54877796274059498683e+06         \\
        &   3 & -5.25000000000000000000e+00   &    &   9 & 2.60000000000000000000e+01         &    &   6 & -8.33349567427633097395e+05        \\
        &   4 & -6.00000000000000000000e+00   &    &  10 & -0.00000000000000000000e+00        &    &   7 & -4.23806772828293405473e+06        \\
        &   5 & -0.00000000000000000000e+00   &    &     &                                    &    &   8 & 3.58496589470378821716e+05         \\
        &     &                               & 11 &   0 & -1.00000000000000000000e+00        &    &   9 & 1.49872407836358295754e+06         \\
      6 &   0 & 1.00000000000000000000e+00    &    &   1 & -1.25500000000000000000e+02        &    &  10 & 5.21977267719965777360e+05         \\
        &   1 & 1.80000000000000000000e+01    &    &   2 & -2.57100000000000000000e+03        &    &  11 & 7.00816817626952833962e+04         \\
        &   2 & 2.44999999999999893419e+01    &    &   3 & -1.59531196289062500000e+04        &    &  12 & 3.56925000000000000000e+03         \\
        &   3 & -1.21875000000000000000e+02   &    &   4 & -2.17014048374422091001e+04        &    &  13 & 5.10000000000000000000e+01         \\
        &   4 & 4.07500000000000000000e+01    &    &   5 & 4.88961204626596372691e+04         &    &  14 & -0.00000000000000000000e+00        \\
        &   5 & 9.00000000000000000000e+00    &    &   6 & 2.47395297945582315151e+04         &    &     &                                    \\
        &   6 & -0.00000000000000000000e+00   &    &   7 & -2.65267486447097398923e+04        & 15 &   0 & -1.00000000000000000000e+00        \\
        &     &                               &    &   8 & -1.06594931640625000000e+04        &    &   1 & -3.16500000000000000000e+02        \\
      7 &   0 & -1.00000000000000000000e+00   &    &   9 & -1.14000000000000000000e+03        &    &   2 & -1.51542500000000000000e+04        \\
        &   1 & -3.05000000000000000000e+01   &    &  10 & -3.15000000000000000000e+01        &    &   3 & -2.55901594619751005666e+05        \\
        &   2 & -5.25000000000000000000e+00   &    &  11 & -0.00000000000000000000e+00        &    &   4 & -1.81766897669971594587e+06        \\
        &   3 & 1.83769531250000000000e+02    &    &     &                                    &    &   5 & -5.47577552606320381165e+06        \\
        &   4 & 2.11445312500000213163e+01    & 12 &   0 & 1.00000000000000000000e+00         &    &   6 & -3.39287989124816888943e+06        \\
        &   5 & -1.11250000000000000000e+02   &    &   1 & 1.63000000000000000000e+02         &    &   7 & 1.32761055815006103367e+07         \\
        &   6 & -1.25000000000000000000e+01   &    &   2 & 4.26500000000000000000e+03         &    &   8 & 8.67954126374729350209e+06         \\
        &   7 & -0.00000000000000000000e+00   &    &   3 & 3.62502430419921875000e+04         &    &   9 & -5.72495353703939169645e+06        \\
        &     &                               &    &   4 & 9.72816786722543474752e+04         &    &  10 & -4.44897483104257006198e+06        \\
      8 &   0 & 1.00000000000000000000e+00    &    &   5 & -2.92123691227543495188e+04        &    &  11 & -1.10378333697890699841e+06        \\
        &   1 & 4.70000000000000000000e+01    &    &   6 & -2.79344760523458826356e+05        &    &  12 & -1.17472772171020405949e+05        \\
        &   2 & 3.47500000000000000000e+02    &    &   7 & 4.62104999846149221412e+04         &    &  13 & -4.93375000000000000000e+03        \\
        &   3 & 1.52800781250000198952e+02    &    &   8 & 8.62397669988747074967e+04         &    &  14 & -5.85000000000000000000e+01        \\
        &   4 & -1.45959277343750000000e+03   &    &   9 & 2.14371234130859302240e+04         &    &  15 & -0.00000000000000000000e+00        \\
        &   5 & 4.47820312500000113687e+02    &    &  10 & 1.72550000000000000000e+03         &    &     &                                    \\
        &   6 & 2.33250000000000000000e+02    &    &  11 &
        3.75000000000000000000e+01         &    &     &   continued in Tab.~\ref{table-AB-2}
        \\

      \end{tabular}
    \end{ruledtabular}
  \end{table*}

\begin{table*}
  \caption{
    Series for the ultimate survival probability, $P_{\infty} =
    \sum_{n=16}^{N} \sum_{m=0}^n c_{nm} \mu_A^{n-m} \mu_B^m$, for the
    heterogeneous lattice $AB$ starting from a single occupied
    $A$-site up to order $N=24$ (continued from Tab.~\ref{table-AB-1}).
    \label{table-AB-2}}
  \begin{ruledtabular}
    \begin{tabular}{ccccccccc}
      $n$ & $m$ & $c_{nm} $ & $n$ & $m$ & $c_{nm}$ &  $n$ & $m$ & $c_{nm} $ \\
      \hline
      16 &   0 & 1.00000000000000000000e+00    &    &   9 & 3.80933065114315605164e+09         &    &   9 & 9.19978211394436798096e+10         \\
      &   1 & 3.83000000000000000000e+02    &    &  10 & 2.79066540292775917053e+09         &    &  10 & -1.41663236622908508301e+11        \\
      &   2 & 2.17725000000000000000e+04    &    &  11 & -1.18839624965687608719e+09        &    &  11 & -3.09544181677179382324e+11        \\
      &   3 & 4.40507505485534784384e+05    &    &  12 & -1.54566718886240291595e+09        &    &  12 & -5.92025291294928512573e+10        \\
      &   4 & 3.88379059831085987389e+06    &    &  13 & -6.04278256410952091217e+08        &    &  13 & 1.17816956489996398926e+11         \\
      &   5 & 1.58357353872441500425e+07    &    &  14 & -1.20264935356224894524e+08        &    &  14 & 9.15751721367468566895e+10         \\
      &   6 & 2.44039435816917717457e+07    &    &  15 & -1.27080503243713695556e+07        &    &  15 & 3.26331674054245605469e+10         \\
      &   7 & -1.85767939421518109739e+07   &    &  16 & -6.68376087843894492835e+05        &    &  16 & 6.70219656522058677673e+09         \\
      &   8 & -6.69576071585644334555e+07   &    &  17 & -1.49485000000000000000e+04        &    &  17 & 8.07064469443650722504e+08         \\
      &   9 & 4.24265176388191117439e+05    &    &  18 & -9.35000000000000000000e+01        &    &  18 & 5.49525110365272164345e+07         \\
      &  10 & 2.54600385214862190187e+07    &    &  19 & -0.00000000000000000000e+00        &    &  19 & 1.94577521464836411178e+06         \\
      &  11 & 1.14637431981598399580e+07    &    &     &                            &    &  20 & 2.97977500000000109139e+04         \\
      &  12 & 2.18359439378216117620e+06    & 20 &   0 & 1.00000000000000000000e+00         &    &  21 & 1.25000000000000000000e+02         \\
      &  13 & 1.89539660865783487679e+05    &    &   1 & 7.39000000000000000000e+02         &    &  22 & -0.00000000000000000000e+00        \\
      &  14 & 6.67674999999999818101e+03    &    &   2 & 7.58940000000000000000e+04         &    &     &                            \\
      &  15 & 6.65000000000000000000e+01    &    &   3 & 2.78482589066410297528e+06         & 23 &   0 & -1.00000000000000000000e+00        \\
      &  16 & -0.00000000000000000000e+00   &    &   4 & 4.77064398069097101688e+07         &    &   1 & -1.11450000000000000000e+03        \\
      &     &                               &    &   5 & 4.30921612129856288433e+08         &    &   2 & -1.66033750000000000000e+05        \\
      17 &   0 & -1.00000000000000000000e+00   &    &   6 & 2.15946993785786914825e+09         &    &   3 & -8.73441611697400361300e+06        \\
      &   1 & -4.58000000000000000000e+02   &    &   7 & 5.80407028293784332275e+09         &    &   4 & -2.18288062553152590990e+08        \\
      &   2 & -3.05770000000000000000e+04   &    &   8 & 5.94638918469835090637e+09         &    &   5 & -2.99289542933531618118e+09        \\
      &   3 & -7.29997349998474353924e+05   &    &   9 & -7.40305007272274589539e+09        &    &   6 & -2.41184413851611099243e+10        \\
      &   4 & -7.78990695947526767850e+06   &    &  10 & -1.82033732679653511047e+10        &    &   7 & -1.17519830906608703613e+11        \\
      &   5 & -4.04899793576145172119e+07   &    &  11 & -2.40532317192275094986e+09        &    &   8 & -3.41927873040872924805e+11        \\
      &   6 & -9.69540386525691747665e+07   &    &  12 & 7.10905857967732429504e+09         &    &   9 & -5.02858276294200012207e+11        \\
      &   7 & -3.89085116880342364311e+07   &    &  13 & 4.71814863830389404297e+09         &    &  10 & 3.15742561692385902405e+10         \\
      &   8 & 2.23379967691832602024e+08    &    &  14 & 1.42660811422320199013e+09         &    &  11 & 1.14735787437879589844e+12         \\
      &   9 & 1.55998095356879800558e+08    &    &  15 & 2.35853332781435012817e+08         &    &  12 & 8.97016221831120605469e+11         \\
      &  10 & -8.29453441516727209091e+07   &    &  16 & 2.12726885021208114922e+07         &    &  13 & -2.27905936943680297852e+11        \\
      &  11 & -8.40138382690373063087e+07   &    &  17 & 9.71620832698821439408e+05         &    &  14 & -5.02435627291505371094e+11        \\
      &  12 & -2.68378383641524799168e+07   &    &  18 & 1.90230000000000000000e+04         &    &  15 & -2.66655503063110687256e+11        \\
      &  13 & -4.09565602203018311411e+06   &    &  19 & 1.03500000000000000000e+02         &    &  16 & -7.74699632413517761230e+10        \\
      &  14 & -2.96166594512939278502e+05   &    &  20 & -0.00000000000000000000e+00        &    &  17 & -1.35731640547505397797e+10        \\
      &  15 & -8.87100000000000000000e+03   &    &     &                            &    &  18 & -1.42167162412259006500e+09        \\
      &  16 & -7.50000000000000000000e+01   & 21 &   0 & -1.00000000000000000000e+00        &    &  19 & -8.53138949911509156227e+07        \\
      &  17 & -0.00000000000000000000e+00   &    &   1 & -8.53000000000000000000e+02        &    &  20 & -2.69011596571039920673e+06        \\
      &     &                               &    &   2 & -9.97227500000000000000e+04        &    &  21 & -3.67522500000000072760e+04        \\
      18 &   0 & 1.00000000000000000000e+00    &    &   3 & -4.15225493661523098126e+06        &    &  22 & -1.36500000000000000000e+02        \\
      &   1 & 5.42000000000000000000e+02    &    &   4 & -8.13538837860736250877e+07        &    &  23 & -0.00000000000000000000e+00        \\
      &   2 & 4.21000000000000000000e+04    &    &   5 & -8.53730911617448449135e+08        &    &     &                            \\
      &   3 & 1.17137547012138389982e+06    &    &   6 & -5.09422722262319755554e+09        & 24 &   0 & 1.00000000000000000000e+00         \\
      &   4 & 1.48443261883842591196e+07    &    &   7 & -1.72895074425995292664e+10        &    &   1 & 1.26300000000000000000e+03         \\
      &   5 & 9.48619801399879902601e+07    &    &   8 & -2.92424105203240814209e+10        &    &   2 & 2.10837500000000000000e+05         \\
      &   6 & 3.07316159384340524673e+08    &    &   9 & -2.19503462337732887268e+09        &    &   3 & 1.23666210482723508030e+07         \\
      &   7 & 3.81732671278393089771e+08    &    &  10 & 6.57535642176373291016e+10         &    &   4 & 3.45435577848057389259e+08         \\
      &   8 & -3.78560457913966596127e+08   &    &  11 & 4.99555141215952529907e+10         &    &   5 & 5.34326366176584434509e+09         \\
      &   9 & -1.09090039428584504128e+09   &    &  12 & -1.67071432757774791718e+10        &    &   6 & 4.91754102809474029541e+10         \\
      &  10 & -7.22351460704629123211e+07   &    &  13 & -2.80141699739296989441e+10        &    &   7 & 2.78581107827627380371e+11         \\
      &  11 & 4.26837997859395503998e+08    &    &  14 & -1.29120452738848495483e+10        &    &   8 & 9.79218318459290283203e+11         \\
      &  12 & 2.37199959095923602581e+08    &    &  15 & -3.17061061754675102234e+09        &    &   9 & 1.98256718408773803711e+12         \\
      &  13 & 5.84676568044034391642e+07    &    &  16 & -4.44081987989710092545e+08        &    &  10 & 1.41222325451664990234e+12         \\
      &  14 & 7.35058582342192064971e+06    &    &  17 & -3.46190648118626475334e+07        &    &  11 & -2.67964226093419287109e+12        \\
      &  15 & 4.50249120122909313068e+05    &    &  18 & -1.38645204595112707466e+06        &    &  12 & -5.34463787449065527344e+12        \\
      &  16 & 1.15980000000000000000e+04    &    &  19 & -2.39322500000000000000e+04        &    &  13 & -1.30437844961177197266e+12        \\
      &  17 & 8.40000000000000000000e+01    &    &  20 & -1.14000000000000000000e+02        &    &  14 & 1.94623313797743408203e+12         \\
      &  18 & -0.00000000000000000000e+00   &    &  21 & -0.00000000000000000000e+00        &    &  15 & 1.74636441085664794922e+12         \\
      &     &                               &    &     &                            &    &  16 & 7.14760408659765625000e+11         \\
      19 &   0 & -1.00000000000000000000e+00   & 22 &   0 & 1.00000000000000000000e+00         &    &  17 & 1.74634440975768798828e+11         \\
      &   1 & -6.35500000000000000000e+02   &    &   1 & 9.78000000000000000000e+02         &    &  18 & 2.64786405199085998535e+10         \\
      &   2 & -5.69645000000000000000e+04   &    &   2 & 1.29406500000000000000e+05         &    &  19 & 2.43568212742165803909e+09         \\
      &   3 & -1.82815355796528002247e+06   &    &   3 & 6.07409790126359928399e+06         &    &  20 & 1.29837631260614901781e+08         \\
      &   4 & -2.71021273926327489316e+07   &    &   4 & 1.34919942776649802923e+08         &    &  21 & 3.66895718129834206775e+06         \\
      &   5 & -2.07776329672791510820e+08   &    &   5 & 1.62617631624924206734e+09         &    &  22 & 4.49402500000000072760e+04         \\
      &   6 & -8.53126758990889191628e+08   &    &   6 & 1.13523417998541107178e+10         &    &  23 & 1.48500000000000000000e+02         \\
      &   7 & -1.69019990104488110542e+09   &    &   7 & 4.67520642304774093628e+10         &    &  24 & -0.00000000000000000000e+00        \\
      &   8 & -3.82997793883490979671e+08   &    &   8 & 1.07905537383186004639e+11         &    &     &                            \\
    \end{tabular}
  \end{ruledtabular}
\end{table*}

%%%%%%%%%%%%%%%%%%%%%%%%%%%%%%%%%%%%%%%%%%%%%

\begin{table*}
  \caption{Series for the configurationally-averaged ultimate survival
    probability, $\langle P_{\infty} \rangle =
    1- \sum_{n=1}^{N} \sum_{m=0}^n \langle c_{nm} \rangle \mu_A^{n-m} \mu_B^m$, for the
    disordered lattice with impurity concentration $p=0.5$ up to order
    $N = 12$.
    \label{table-p0.5}
  }
  \begin{ruledtabular}
    \begin{tabular}{ccccccccc}
      \emph{n} & \emph{m} &$\langle c_{nm} \rangle$ &\emph{n} & \emph{m} & $\langle c_{nm} \rangle$ &
      \emph{n} & \emph{m} & $\langle c_{nm} \rangle$ \\
      \hline
      0& 0& 1.00000000000000000000e+00   & 7 & 0& -1.02856445312500000000e+00    &   & 4& 8.57398099693330948412e+01 \\
       &  &                              &   & 1& 6.71752929687499911182e+00     &   & 5&-1.59956149689860012586e+03 \\
      1& 0& -5.00000000000000000000e-01  &   & 2& -7.88159179687500000000e+00    &   & 6& 8.57398099692479576106e+01 \\
       & 1& -5.00000000000000000000e-01  &   & 3& -3.63503417968750000000e+01    &   & 7& -3.78991681481602029180e+02\\
       &  &                              &  & 4& -3.63503417968750000000e+01    &   & 8& 3.15666996812323645827e+02 \\
      2& 0& 0.00000000000000000000e+00   &  & 5& -7.88159179687500000000e+00    &   & 9& -1.12182894054082581192e+02 \\
   & 1& -1.00000000000000000000e+00     &   & 6& 6.71752929687500000000e+00     &   &10& 1.97285068645925996123e+01 \\
   & 2& 0.00000000000000000000e+00      &   & 7& -1.02856445312500000000e+00    &       &                             \\
   &  &                                 &   &  &                                &11&  0& -6.38827981592922071741e+01  \\
 3&  0& 1.25000000000000000000e-01      & 8&  0& 1.91795349121094105271e+00     &   & 1& 2.76683537172620788169e+02   \\ 
   & 1& -1.12500000000000000000e+00     &   & 1& -1.00450286865234925671e+01    &   & 2& -5.81148325010243183897e+02   \\
   & 2& -1.12500000000000000000e+00     &   & 2& 4.34407348632823939738e+01     &   & 3& 1.08098985786890170857e+03   \\
   & 3& 1.25000000000000000000e-01      &   & 3& -1.05206558227552989138e+02    &   & 4& -1.09115308153965179372e+03   \\
   &  &                                 &   & 4& -7.39358825683648177574e+01    &   & 5& -2.15854719928576787424e+03   \\
4&  0& 2.18750000000000000000e-01      &   & 5& -1.05206558227541165706e+02    &   & 6& -2.15854719928370877824e+03   \\
   & 1& -8.75000000000000000000e-01     &   & 6& 4.34407348632839358515e+01     &   & 7& -1.09115308154012654995e+03   \\
   & 2& 2.18750000000000000000e-01      &   & 7& -1.00450286865224249766e+01    &   & 8& 1.08098985786938828824e+03    \\
   & 3& -8.75000000000000000000e-01     &   & 8& 1.91795349121088110067e+00     &   & 9& -5.81148325010462372120e+02   \\
   & 4& 2.18750000000000000000e-01      &   &   &                               &   &10& 2.76683537172492606260e+02    \\
   &  &                                 & 9&  0& -9.13367027706589240665e+00    &   &11& -6.38827981592518412413e+01   \\
 5&  0& 1.56250000000000000000e-01      &   & 1& 4.07988169988005395794e+01     &       &                              \\
   & 1& 3.43750000000000000000e-01      &   & 2& -5.42188301086439778942e+01    &12&  0& 1.82464000646571463449e+02    \\
   & 2& -6.00000000000000000000e+00     &   & 3& 6.09921527438696244872e+01     &   & 1& -9.07237726925243578080e+02   \\
   & 3& -6.00000000000000000000e+00     &   & 4& -3.40893967946386396761e+02    &   & 2& 2.01989525183668274622e+03    \\
   & 4& 3.43750000000000000000e-01      &   & 5& -3.40893967946387931534e+02    &   & 3& -2.51445338021211500745e+03   \\
   & 5& 1.56250000000000000000e-01      &   & 6& 6.09921527438643025221e+01     &   & 4& 4.03956355580561148599e+03    \\
   &  &                                 &   & 7& -5.42188301086414981000e+01    &   & 5& -8.47482562600732853753e+03   \\
6  & 0& 3.45703125000000000000e-01      &   & 8& 4.07988169988003761546e+01     &   & 6& -3.63662968564300990693e+03   \\
   & 1& -1.91406250000000000000e-01     &   & 9& -9.13367027706600786985e+00    &   & 7& -8.47482562600957498944e+03   \\
   & 2& -1.76757812500000000000e+00     &   &  &                                &   & 8& 4.03956355578774446258e+03    \\
   & 3& -2.53984375000000000000e+01     &10&  0& 1.97285068645884216210e+01     &   & 9& -2.51445338020943927404e+03   \\
   & 4& -1.76757812500000000000e+00     &   & 1& -1.12182894054052439969e+02    &   &10& 2.01989525183534055941e+03    \\
   & 5& 1.56250000000000000000e-01      &   & 2& 3.15666996812357751878e+02     &   &11& -9.07237726928176357433e+02   \\
   & 6& 3.45703125000000000000e-01      &   & 3& -3.78991681481601744963e+02    &   &12& 1.82464000646704874953e+02    \\
    \end{tabular}
  \end{ruledtabular}
\end{table*}

%%%%%%%%%%%%%%%%%%%%%%%%%%%%%%%%%%%%%%%%%%%%%%%%%%%%%%%%%%%%%%%%%%%%%%%%%%%5

%\bibliographystyle{apsrev}%{unsrt} % Choose Phys. Rev. style for bibliography
%\bibliography{rainbology} 

\end{document}